\documentclass[twocolumn,showpacs,nofootinbib,aps,superscriptaddress,prd]{revtex4}
\usepackage{graphicx,dcolumn,bm,amsfonts,amsmath,color,xcolor,amsthm}
\usepackage[dvipdfm,colorlinks=true,citecolor=blue, linkcolor=blue,urlcolor=blue]{hyperref}
\usepackage{slashed}

\allowdisplaybreaks
\begin{document}
\title{$K_0^\ast(1430)$  Twist-2 Distribution Amplitude and $B_s,D_s \to K_0^\ast(1430)$ Transition Form Factors}
\author{Dong Huang}
\address{School of Physics and Mechatronic Engineering, Guizhou Minzu University, Guiyang 550025, P.R. China}
\author{Tao Zhong\footnote{Corresponding author}}
\email{zhongtao1219@sina.com}
\address{School of Physics and Mechatronic Engineering, Guizhou Minzu University, Guiyang 550025, P.R. China}
\author{Hai-Bing Fu}
\email{fuhb@cqu.edu.cn}
\address{School of Physics and Mechatronic Engineering, Guizhou Minzu University, Guiyang 550025, P.R. China}
\address{Department of Physics, Chongqing University, Chongqing 401331, P.R. China}
\author{Zai-Hui Wu}
\address{School of Physics and Mechatronic Engineering, Guizhou Minzu University, Guiyang 550025, P.R. China}
\author{Xing-Gang Wu}
\email{wuxg@cqu.edu.cn}
\address{Department of Physics, Chongqing University, Chongqing 401331, P.R. China}
\author{Hong Tong}
\address{School of Physics and Mechatronic Engineering, Guizhou Minzu University, Guiyang 550025, P.R. China}
\date{\today}

\begin{abstract}
Based on the scenario that the $K_0^\ast(1430)$ is viewed as the ground state of $s\bar{q}$ or $q\bar{s}$, we study the $K_0^\ast(1430)$ leading-twist distribution amplitude (DA) $\phi_{2;K_0^\ast}(x,\mu)$ with the QCD sum rules in the framework of background field theory. A more reasonable sum rule formula for $\xi$-moments $\langle\xi^n\rangle_{2;K_0^\ast}$ is suggested, which eliminates the influence brought by the fact that the sum rule of $\langle\xi^0_p\rangle_{3;K_0^\ast}$ cannot be normalized in whole Borel region. More accurate values of the first ten $\xi$-moments, $\langle\xi^n\rangle_{2;K_0^\ast} (n = 1,2,\cdots,10)$, are evaluated. A new light-cone harmonic oscillator (LCHO) model for $K_0^\ast(1430)$ leading-twist DA is established for the first times. By fitting the resulted values of $\langle\xi^n\rangle_{2;K_0^\ast} (n = 1,2,\cdots,10)$ via the least squares method, the behavior of $K_0^\ast(1430)$ leading-twist DA described with LCHO model is determined. Further, by adopting the light-cone QCD sum rules, we calculate the $B_s,D_s \to K_0^\ast(1430)$ transition form factors and branching fractions of the semileptonic decays $B_s,D_s \to K_0^\ast(1430) \ell \nu_\ell$. The corresponding numerical results can be used to extract the Cabibbo-Kobayashi-Maskawa matrix elements by combining the relative experimental data in the future.
\end{abstract}

\maketitle

\section{INTRODUCTION}

Currently, the exclusive determinations of the Cabibbo-Kobayashi-Maskawa (CKM) matrix elements $|V_{ub}|$ and $|V_{cd}|$ are mainly dominated by the semileptonic $B\to\pi\ell\bar{\nu}$ and $D\to\pi\ell\nu$ decays, respectively~\cite{Workman:2022ynf}. The study of the semileptonic $B_s, D_s \to K_0^\ast(1430) \ell \nu_\ell$ decays can provide a new choice and supplement for extracting these two matrix elements.

In the theoretical studies on the semileptonic $B_s, D_s \to K_0^\ast(1430)$ decays, the most challenging parts are the calculations of the transition form factors (TFFs). These TFFs are mainly dominated by the short-distance dynamics in the large recoil region and the soft dynamics in the region of small recoil. Although the momentum dependence of the $B_s\to K_0^\ast(1430)$ TFFs calculated by Faustov and Galkin in the relativistic quark model (RQM) is determined in the whole accessible kinematical range~\cite{Faustov:2013ima}, the commonly used research methods such as perterbative QCD (pQCD) factorization, QCD sum rules (QCDSRs), light-cone QCD sum rules (LCSRs) and Lattice QCD (LQCD) theory, etc., are usually only applicable to calculate the TFFs in specific $q^2$ regions. The corresponding results should be extrapolated to the whole kinematic region by adopting appropriate parametrization form. Usually, the LQCD works well in the small recoil region. However, there are currently no LQCD calculations of the $B_s, D_s \to K_0^\ast(1430)$ TFFs in the literature. The pQCD factorization method is usually applicable near the large recoil point, and which has been used to calculate $B_s \to K_0^\ast(1430)$ TFFs in Refs.~\cite{Li:2008tk, Zhang:2010af, Chen:2021oul}. The QCDSR and LCSR estimations for TFFs are effective in low and intermediate $q^2$ regions. In Ref.~\cite{Yang:2005bv}, the TFFs $f_+^{B_s\to K_0^\ast}(q^2)$ and $f_+^{D_s\to K_0^\ast}(q^2)$ are calculated by adopting the QCDSRs with three-point correlation function (correlator). In which, the TFFs are parameterized as vacuum condensates with different dimension. The three-point QCDSRs is then used to calculate TFFs $f^{B_s\to K_0^\ast}_{\pm,T}(q^2)$ and further analyze the rare semileptonic $B_s\to K_0^\ast(1430) \ell^+ \ell^-$ decays~\cite{Ghahramany:2009zz}. More researches on the $B_s\to K_0^\ast(1430)$ TFFs are performed in the framework of LCSRs. Different from the three-point QCDSRs, the TFFs calculated within LCSRs are parameterized as the initial or final meson distribution amplitudes (DAs) arranged with different twist structures. Starting from different correlator, the LCSRs of $B_s\to K_0^\ast(1430)$ TFFs are expressed as the convolution integrals of the $B_s$ meson DAs~\cite{Khosravi:2022fzo}, the $K_0^\ast(1430)$ twist-2, 3 DAs~\cite{Wang:2008da, Wang:2014vra, Wang:2014upa}, only $K_0^\ast(1430)$ twist-2 DA~\cite{Sun:2010nv} or twist-3 DAs~\cite{Han:2013zg}, respectively. In addition, there are some other studies on the $D_s \to K_0^*(1430)$ TFFs in the literature. For example, Ref.~\cite{Cheng:2002ai} extracts the TFF $f_+^{D_s\to K_0^\ast}(0)$ from the data of the hadronic $D$ decay in the generalized factorization model.

In this paper, we will calculate the $B_s, D_s \to K_0^\ast(1430)$ TFFs within the LCSR method by adopting the traditional current correlator. Specifically, we will adopt the calculation technology for the operator product expansion (OPE) in Refs.~\cite{Duplancic:2008ix, Duplancic:2008tk}. That is, the usual suppression by the powers of the Borel parameter for the higher twist contributions is transferred as exponential suppression through the integration by parts.

Usually, the LCSRs of the $B_s, D_s \to K_0^\ast(1430)$ TFFs are dominated by the contributions proportional to the $K_0^\ast(1430)$ leading-twist DA $\phi_{2;K_0^\ast}(x,\mu)$. Then $\phi_{2;K_0^\ast}(x,\mu)$ is the mainly error source of these TFFs. In order to obtain accurate predictions for the semileptonic $B_s, D_s \to K_0^*(1430)$ decays, one should determine the accurate behavior of the $K_0^\ast(1430)$ leading-twist DA. Before this, however, we should clarify our understanding of quark content of scalar $K_0^\ast(1430)$ meson. Different from the scalar mesons below $1\ {\rm GeV}$ $-$ for which there is no general agreement on the multiple candidates such as conventional $q\bar{q}$ states~\cite{Cheng:2005nb}, meson-meson molecular states~\cite{Weinstein:1982gc}, tetraquark states~\cite{Jaffe:1976ig, Jaffe:1976ih, Wang:2010pn} and so on $-$ the $K_0^\ast(1430)$ is predominantly viewed as the $s\bar{q}$ or $q\bar{s}$ state in almost all the relative researches. The only controversy about $K_0^\ast(1430)$ lies between the following two scenarios: Scenario 1 (S1), the $K_0^\ast(1430)$ is assumed to be the excited state corresponding to the ground state below $1\ {\rm GeV}$; Scenario 2 (S2), the $K_0^\ast(1430)$ is viewed as the ground state while the nonet scalars below $1\ {\rm GeV}$ may be considered as four-quark bound states. Ref.~\cite{Du:2004ki} calculates the masses and decay constants of $I=1/2$ scalar mesons with QCDSRs. Its result favors that $K_0^\ast(1430)$ is the lowest scalar stats of $s\bar{q}$ or $q\bar{s}$, i.e., S2. Later, based on S2, Refs.~\cite{Lu:2006fr, Han:2013zg} study the $K_0^\ast(1430)$ twist-3 DAs with QCDSRs. Besides, the research on the two-body decays of $B_{(s)}$ containing $K_0^\ast(1430)$ in Ref.~\cite{Chen:2021oul} also supports that $K_0^\ast(1430)$ should be described as the lowest-lying p-wave state rather than the first excited one. Therefore, our research work on $K_0^\ast(1430)$ in this paper will also take S2 as the starting point. In numerical analysis, only the data corresponding to S2 in literature will be adopted for consistency.

The $\phi_{2;K_0^\ast}(x,\mu)$ has been investigated in Ref.~\cite{Cheng:2005nb} with QCDSRs, and in Ref.~\cite{Chen:2021oul} with light-front (LF) approach. In Ref.~\cite{Cheng:2005nb}, the first two nonzero $\xi$-moments and Gegenbauer moments are calculated and substituted into the truncation form of the Gegenbauer expansion series (TF model) of $\phi_{2;K_0^\ast}(x,\mu)$ to predict the behavior of the $K_0^\ast(1430)$ leading-twist DA. However, our recent analysis on the commonly used phenomenological model of the pionic leading-twist DA shows that the simple TF model is far from sufficient to describe the accurate behavior of DA~\cite{Zhong:2022lmn}. Therefore, we will re study the $K_0^\ast(1430)$ leading-twist DA with the QCDSRs in the framework of background field theory (BFT)~\cite{Huang:1989gv}. In order to obtain more accurate behavior of $\phi_{2;K_0^\ast}(u,\mu)$, a research scheme suggested in our previous study on the pionic leading-twist DA~\cite{Zhong:2021epq} will be adopted. This scheme has been used to researches on the kaon leading-twist DA~\cite{Zhong:2022ecl} and the $a_1(1260)$ meson longitudinal twist-2 DA~\cite{Hu:2021lkl}. Specifically, we will construct a light-cone harmonic oscillator (LCHO) model based on the Brodsky-Huang-Lepage (BHL) description~\cite{BHL} for $\phi_{2;K_0^\ast}(u,\mu)$. The model parameters will be determined by fitting the first ten $\xi$-moments with the least squares method. More accurate values of those $\xi$-moments will be evaluated by using a new sum rule formula suggested in Ref.~\cite{Zhong:2021epq}.

This paper is organized as follows. In Sec. II, the sum rules for the $\xi$-moments of the $K_0^\ast(1430)$ leading-twist DA and the LCSRs of the $B_s, D_s \to K_0^\ast(1430)$ TFFs $f_{\pm,T}(q^2)$ are derived, a new LCHO model for the $K_0^\ast(1430)$ leading-twist DA is established at the first time. In Sec. III, we provide the relevant numerical results. The final section is reserved for a summary. As a by-product, the numerical calculations for the $K_0^\ast(1430)$ leading-twist DA and $B_s, D_s \to K_0^\ast(1430)$ TFFs and branching fractions corresponding to S1 are performed, and shown in Appendix B.

\section{THEORETICAL FRAMEWORK}

\subsection{Sum rules for the $\xi$-moments of $\phi_{2;K_0^\ast}(x,\mu)$}\label{Sec:IIA}

The leading-twist DA $\phi_{2;K_0^\ast}(x,\mu)$ of the scalar $K_0^\ast(1430)^+$ meson with quark content $u\bar{s}$ are given by~\cite{Cheng:2005nb}
\begin{align}
&\langle 0\left| \bar{s}(z)\gamma_\mu u(-z) \right|K_0^{\ast +}\rangle \nonumber\\
&\quad\quad\quad\quad = \bar{f}_{K_0^\ast} p_\mu \int^1_0 du e^{i(2u-1)p\cdot z} \phi_{2;K_0^\ast}(u,\mu),
\label{T2DA}
\end{align}
where $z^2 = 0$, $\bar{f}_{K_0^\ast}$ and $m_{K_0^\ast}$ are the decay constant and mass of $K_0^\ast(1430)$ meson, respectively. Expanding both sides of Eq.~\eqref{T2DA} into series of $z$, one can get
\begin{align}
\langle 0| \bar{s}(0)\gamma_\mu (iz\cdot \tensor{D})^n u(0) |K_0^{\ast +}\rangle = p_\mu (p\cdot z)^n \bar{f}_{K_0^\ast} \langle\xi^n\rangle_{2;K_0^\ast},
\label{XinT2DAMatrixElement}
\end{align}
where ${D_\mu } = {\partial_\mu } - i{g_s}{T^A}\mathcal{A}_{\mu}^{A}(x)(A=1,...,8)$ is the fundamental representation of the gauge covariant derivative, and
\begin{align}
\langle\xi^n\rangle_{2;K_0^\ast} = \int^1_0 du (2u-1)^n \phi_{2;K_0^\ast}(u).
\label{xinT2DA}
\end{align}
is the $n$th $\xi$-moment. Then, in order to derive the sum rules of the $\xi$-moments $\langle{\xi^{n}}\rangle_{2;K_0^\ast}$, we introduce the following correlator,
\begin{align}
\Pi_{2;K_0^\ast}(z,q) &= i\int d^4x {e^{iq\cdot x}}\langle 0|T\{ {J_n}(x),\hat{J}_0^\dag (0)\} |0\rangle
\nonumber\\
&= {(z\cdot q)^{n+1}}{I_{2;K_0^\ast}}(q^2)
\label{correlatorDA}
\end{align}
with the interpolating currents
\begin{align}
J_n(x) &= \bar{s}(x)\slashed{z}(iz\cdot\tensor{D})^n u(x),
\nonumber\\
\hat{J}_0^\dagger(0) &= \bar{u}(0) s(0).
\end{align}

We now perform the OPE for the correlator~\eqref{correlatorDA} in the deep Euclidean region. The calculation is carried out in the framework of BFT~\cite{Huang:1989gv}. By decomposing quark and gluon fields into classical background fields describing nonperturbative effects and quantum fields describing perturbative effects, BFT can provide clear physical images for the separation of long- and short-range dynamics in OPE. Based on the basic assumptions and Feynman rules of BFT~\cite{Huang:1989gv}, the correlator~\eqref{correlatorDA} can be rewritten as
\begin{align}
{\Pi_{2;K_0^\ast}}(z\cdot q) &= i\int {{d^4}x{e^{iq \cdot x}}}
\nonumber\\
&\times \Big\{-{\rm Tr} \langle 0|S_F^s(0,x) \slashed{z} (iz\cdot\tensor{D})^n S_F^u(x,0)|0\rangle \nonumber\\
&+ {\rm Tr} \langle 0|\bar{s}(x)s(0) \slashed{z} (iz\cdot\tensor{D})^n S_F^u(x,0)|0\rangle \nonumber\\
&+ {\rm Tr} \langle 0|S_F^s(0,x) \slashed{z} (iz\cdot\tensor{D})^n \bar{u}(0)u(x)|0\rangle \nonumber\\
&+ \cdots \Big\},
\label{correlatorDAExpansion}
\end{align}
where $\rm{Tr}$ indicates trace for the $\gamma$-matrix and color matrix, $S_F^s(0,x)$ indicate the $s$-quark propagator from $x$ to $0$, $S_F^u(x,0)$ stands for the $u$-quark propagator from $0$ to $x$, $\slashed{z} (iz\cdot\tensor{D})^n$ are the vertex operators from current $J_n(x)$, respectively. The expressions up to dimension-six of the quark propagator, the vertex operator, and the vacuum matrix elements such as $\langle 0|\bar{s}(x)s(0)\rangle$ and $\langle 0|\bar{u}(0)u(x)\rangle$ have been derived and given in Refs.~\cite{Zhong:2011rg, Zhong:2014jla, Zhong:2021epq, Hu:2021zmy}. By substituting those corresponding formula into Eq.~\eqref{correlatorDAExpansion}, the OPE of correlator~\eqref{correlatorDA}, $I_{2;K_0^\ast}^{\rm qcd}(q^2)$, can be obtained.

By inserting a complete set of hadronic states into correlator~\eqref{correlatorDA} in physical region, whose hadronic representation can be read as
\begin{align}
{\rm Im} I_{2;K_0^\ast}^{\rm had}(s) &= \pi m_{K_0^\ast} \delta(s - m_{K_0^\ast}^2) \bar{f}_{K_0^\ast}^2 \langle\xi^n\rangle_{2;K_0^\ast} \langle\xi^0_p\rangle_{3;K_0^\ast} \nonumber\\
&+ {\rm Im} I_{2;K_0^\ast}^{\rm pert}(s) \theta(s - s_{K_0^\ast}),
\label{HadronicRepresentation}
\end{align}
where $s_{K_0^\ast}$ is the continuum threshold, $\langle\xi^0_p\rangle_{3;K_0^\ast}$ is the zeroth $\xi$-moment of $K_0^\ast(1430)^+$ two-particle twist-3 DA $\phi_{3;K_0^\ast}^p(u,\mu)$. In the calculation of Eq.~\eqref{HadronicRepresentation}, the matrix element formula in Eq.~\eqref{XinT2DAMatrixElement}, $\langle K_0^{\ast +}(p)| \hat{J}_0^\dagger(0) |0\rangle = \bar{f}_{K_0^\ast} m_{K_0^\ast} \langle\xi^0_p\rangle_{3;K_0^\ast}$ and the quark-hadron duality have been used.

Substituting the resulted OPE and hadronic representation of correlator~\eqref{correlatorDA}, i.e., $I_{2;K_0^\ast}^{\rm qcd}(q^2)$ and ${\rm Im} I_{2;K_0^\ast}^{\rm had}(s)$, into the following dispersion relation after Borel transformation,
\begin{align}
\frac{1}{\pi}\frac{1}{M^2} \int_{m_s^2} ds e^{-s/M^2} {\rm Im} I_{2;K_0^\ast}^{\rm had}(s) = {\hat L}_M I_{2;K_0^\ast}^{\rm qcd}(q^2),
\end{align}
the sum rules of $\langle\xi^n\rangle_{2;K_0^\ast} \times \langle\xi^n_p\rangle_{3;K_0^\ast}$ reads:
\begin{widetext}
\begin{align}
&\frac{\langle\xi^n\rangle_{2;K_0^\ast} \langle\xi^0_p\rangle_{3;K_0^\ast} m_{K_0^\ast} \bar{f}^2_{K_0^\ast}}{M^2 e^{m_{K_0^\ast}^2/M^2}} = \frac{1}{\pi} \frac{1}{M^2} \int^{s_{K_0^\ast}}_{m_s^2} ds e^{-s/M^2} {\rm Im} I_{2;K_0^\ast}^{\rm pert}(s) + \Big( 1 + \frac{m_s m_u}{2M^2} + \frac{2n+1}{2} \frac{m_s^2}{M^2} \Big) \frac{\langle\bar{s}s\rangle}{M^2} + \Big( -1 - \frac{m_s m_u}{2M^2} \nonumber\\
&+ \frac{m_s^2}{M^2} \Big) \frac{(-1)^n \langle\bar{u}u\rangle}{M^2} + \hat{I}_{\langle G^2\rangle}(M^2) + \hat{I}_{\langle G^2\rangle}^{m_s^3}(M^2) + \Big( -\frac{2n}{3} - \frac{8n-9}{36} \frac{m_s m_u}{M^2} \Big) \frac{\langle g_s\bar{s}\sigma TGs\rangle}{(M^2)^2} + \Big[ \frac{2n}{3} \Big( 1 - \frac{m_s^2}{M^2} \Big) + \frac{8n-9}{36} \frac{m_s m_u}{M^2} \nonumber\\
&+ \frac{m_s^2}{4M^2} \Big] \frac{(-1)^n \langle g_s\bar{u}\sigma TGu\rangle}{(M^2)^2} + \frac{2(n+3)}{81}m_u \frac{\langle g_s\bar{s}s\rangle^2}{(M^2)^3} + \frac{-2(n+3)}{81} m_s \Big( 1 - \frac{m_s^2}{M^2} \Big) \frac{(-1)^n \langle g_s\bar{u}u\rangle^2}{(M^2)^3} + \hat{I}_{\langle G^3\rangle}(M^2) + \hat{I}_{\langle G^3\rangle}^{m_s^3}(M^2) \nonumber\\
&+ \hat{I}_{\langle q^4\rangle}(M^2) + \hat{I}_{\langle q^4\rangle}^{m_s^3}(M^2),
\label{SRxinxi0}
\end{align}
where the imaginary part of the perturbative contribution
\begin{align}
{\rm Im} I_{2;K_0^\ast}^{\rm pert}(s) &= -\frac{3}{16\pi (n+1)(n+2)} \Big\{ m_s \Big[ \Big( 1 - \frac{2m_s^2}{s} \Big)^{n+1} \Big( 2(n+1) \Big( 1 - \frac{m_s^2}{s} \Big) + 1 \Big) + (-1)^n \Big] \nonumber\\
&- m_u \Big[ \Big( 1 - \frac{2m_s^2}{s} \Big)^{n+1} \Big( -2(n+1) \Big( 1 - \frac{m_s^2}{s} \Big) + 2n + 3 \Big) + (-1)^n(2n+3) \Big] \Big\},
\label{Imxinxi0}
\end{align}
\end{widetext}
$\hat{I}_{\langle G^2\rangle}(M^2)$, $\hat{I}_{\langle G^3\rangle}(M^2)$, $\hat{I}_{\langle q^4\rangle}(M^2)$ are the contributions proportional to the double-gluon condensate $\langle\alpha_s G^2\rangle$, triple-gluon condensate $\langle g_s^3fG^3\rangle$ and four-quark condensate $\langle g_s^2\bar{u}u\rangle^2$, and $\hat{I}^{m_s^3}_{\langle G^2\rangle}(M^2)$, $\hat{I}^{m_s^3}_{\langle G^3\rangle}(M^2)$, $\hat{I}^{m_s^3}_{\langle q^4\rangle}(M^2)$ are corresponding $\mathcal{O}(m_s^3)$-corrections, respectively. The specific expressions for those terms are exhibited in Appendix A for convenience. In Eq.~\eqref{SRxinxi0}, in addition, $M$ is the Borel parameter, $m_u$ and $m_s$ are the current quark masses of $u$ and $s$ quarks, $\langle\bar{u}u\rangle$ and $\langle\bar{s}s\rangle$ are double-quark condensates with $\langle\bar{s}s\rangle / \langle\bar{u}u\rangle = \kappa$, $\langle g_s\bar{u}\sigma TGu\rangle$ and $\langle g_s\bar{s}\sigma TGs\rangle$ are quark-gluon mixed condensates, $\langle g_s\bar{u}u\rangle^2$ and $\langle g_s\bar{s}s\rangle^2$ are four-quark condensates. In the calculation of OPE, the $SU_f(3)$ breaking effect is considered. Specifically, the full $s$ quark mass effect in the perterbative part is preserved; the $s$ quark mass corrections proportional to $m_s^{\le 3}$ for condensate terms are calculated owing to $m_s \sim 0.1{\rm GeV}$, while $m_u^2 \sim 0$ is adopted due to smallness.

In particular, the sum rules~\eqref{SRxinxi0} is regarded as that for $\langle\xi^n\rangle_{2;K_0^\ast} \times \langle\xi^n_p\rangle_{3;K_0^\ast}$ instead of $\langle\xi^n\rangle_{2;K_0^\ast}$ in this work due to dependence of $\langle\xi^n_p\rangle_{3;K_0^\ast}$ on the Borel parameter as suggested in Ref.~\cite{Zhong:2021epq}. This assumption can be confirmed by the sum rule of $\langle\xi^n_p\rangle_{3;K_0^\ast}$ derived from the correlator $i\int d^4x e^{iq\cdot x} \langle 0| \hat{J}_0(0) \hat{J}_0^\dagger(0) |0\rangle$. Following the above sum rule calculation procedure performed for correlator~\eqref{correlatorDA}, one can easy obtain,
\begin{widetext}
\begin{align}
&\frac{m_{K_0^\ast}^2 \bar{f}_{K_0^\ast}^2 \langle\xi^0_p\rangle_{3;K_0^\ast}^2}{M^2 e^{m_{K_0^\ast}^2/M^2}} = \frac{1}{\pi} \frac{1}{M^2} \int^{s_{K_0^\ast}}_{m_s^2} ds e^{-s/M^2} {\rm Im} I^{\rm pert}_{3;K_0^\ast}(s) + \Big( \frac{m_s}{2} + m_u \Big) \frac{\langle\bar{s}s\rangle}{M^2} + \Big( \frac{m_u}{2} + m_s - \frac{m_s^3}{M^2} \Big) \frac{\langle\bar{u}u\rangle}{M^2} + \frac{1}{24\pi} \Big( 3 - \frac{4m_s^2}{M^2} \Big) \nonumber\\
&\times \frac{\langle\alpha_sG^2\rangle}{M^2} + \frac{m_u}{2} \frac{\langle g_s\bar{s}\sigma TGs\rangle}{(M^2)^2} + \frac{m_s}{2} \Big( 1 - \frac{3m_s^2}{2M^2} \Big) \frac{\langle g_s\bar{u}\sigma TGu\rangle}{(M^2)^2} - \frac{4}{27} \frac{\langle g_s\bar{s}s\rangle^2}{(M^2)^2} - \frac{4}{27} \Big( 1 - \frac{5m_s^2}{4M^2} \Big) \frac{\langle g_s\bar{u}u\rangle^2}{(M^2)^2} - \frac{1}{96\pi^2} \frac{m_s^2}{M^2} \frac{\langle g_s^3fG^3\rangle}{(M^2)^2} \nonumber\\
&+ \frac{1}{972\pi^2} \Big\{ -12 \Big( -\ln\frac{M^2}{\mu^2} \Big) + 70 + \frac{m_s^2}{M^2} \Big[ 15 \Big( -\ln\frac{M^2}{\mu^2} \Big) + 93 \Big] \Big\} \frac{(2+\kappa^2) \langle g_s^2\bar{u}u\rangle^2}{(M^2)^2},
\label{SRxi0xi0}
\end{align}
with
\begin{align}
{\rm Im} I^{\rm pert}_{3;K_0^\ast}(s) &= \frac{3s}{8\pi} \Big( 1 - \frac{m_s^2}{s} \Big)^2 \Big[ 3-2\Big( 1 - \frac{m_s^2}{s} \Big) \Big]
-\frac{3}{4\pi} m_s^2\Big( 1 - \frac{m_s^2}{s} \Big)^2.
\end{align}
\end{widetext}
As discussed above and suggested in Ref.~\cite{Zhong:2021epq}, a better sum rules for $\langle\xi^n\rangle_{2;K_0^\ast}$ is suggested as
\begin{align}
\langle\xi^n\rangle_{2;K_0^\ast} = \frac{\Big(\langle\xi^n\rangle_{2;K_0^\ast} \times \langle\xi^0_p\rangle_{3;K_0^\ast} \Big)|_{\rm From\ Eq.~\eqref{SRxinxi0}}}{\sqrt{\langle\xi^0_p\rangle_{3;K_0^\ast}^2} |_{\rm From\ Eq.~\eqref{SRxi0xi0}}}.   \label{SRxin}
\end{align}
It should be noted that, in the numerical calculations about $\langle\xi^n\rangle_{2;K_0^\ast}$ in Sec. III, we take the scale $\mu = M$ as usual. Then the scale dependency of $\langle\xi^n\rangle_{2;K_0^\ast}$ is achieved through the Borel parameter $M$ and the scale dependency of input parameters such as various vacuum condensates, quark masses, $K_0^\ast(1430)$ decay constant, etc.

\subsection{LCHO model for ${\phi _{2;K_0^\ast}}(x,\mu)$ based on BHL prescription}\label{Sec:IIB}

The $K_0^\ast(1430)$ leading-twist DA, ${\phi _{2;K_0^\ast}}(x,\mu)$, describes the momentum fraction distribution of partons in $K_0^\ast(1430)$ meson for the lowest Fock state. The ${\phi _{2;K_0^\ast}}(x,\mu)$ is a universal nonperturbative objects, and which should be studied with nonperturbative QCD. However, we usually can only use the method of combining nonperturbative QCD and phenomenological model to study ${\phi _{2;K_0^\ast}}(x,\mu)$ due to the difficulty of nonperturbative QCD. In Sec.~\ref{Sec:IIA}, we have calculated the $n$th $\xi$-moment with nonperturbative QCDSR method. In this subsection, we will construct a LCHO model to describe the overall behavior of ${\phi _{2;K_0^\ast}}(x,\mu)$ based on BHL prescription~\cite{BHL}.

The starting point of BHL prescription is the assumption that there is a connection between the equal-times wave function (WF) in the rest frame and the light-cone WF. Through this assumption, one can map the approximate bound state solution in the quark model for meson in the rest frame to the light-cone frame by equating the off-shell propagator in the two frames, thus obtaining the LCHO model of the meson WFs~\cite{Huang:1994dy}. The LCHO model has good end point behavior, which is helpful to suppress the end point singularity in certain processes, so as to obtain more reliable theoretical predictions. So far, LCHO model has been widely used in the study of various meson WFs or DAs and has been continuously improved~\cite{Zhong:2014jla, Zhong:2021epq, Zhong:2022ecl, Hu:2021lkl, Huang:1994dy, Cao:1997hw, Huang:2004fn, Wu:2005kq, Huang:2006wt, Wu:2011gf, Wu:2012kw, Huang:2013gra, Huang:2013yya, Zhong:2014fma, Zhong:2015nxa, Zhong:2016kuv, Zhang:2017rwz, Zhong:2018exo, Zhang:2021wnv}. Formally, the WF of $K_0^\ast(1430)$ leading-twist DA can be expressed as
\begin{align}
\Psi_{2;K_0^\ast}(x, \mathbf{k}_\perp) = \chi_{2;K_0^\ast}(x, \mathbf{k}_\perp) \Psi^R_{2;K_0^\ast}(x, \mathbf{k}_\perp),
\label{eq:WF}
\end{align}
where $\mathbf{k}_\perp$ is the transverse momentum, $\chi_{2;K_0^\ast}(x, \mathbf{k}_\perp)$ stands for the spin-space WF coming from the Wigner-Melosh rotation. As a scalar meson, the spin WF of $K_0^\ast(1430)$ should be~\cite{Zhong:2022ecl}
\begin{align}
\chi_{2;K_0^\ast}(x, \mathbf{k}_\perp) = \frac{\widetilde{m}}{\sqrt{\mathbf{k}_\perp^2 + \widetilde{m}^2}},
\label{eq:spinWF}
\end{align}
where $\widetilde{m} = \hat{m}_q x + \hat{m}_s\bar{x}$ with $q = u/d$ and $\bar{x} = 1-x$. $\hat{m}_q$ and $\hat{m}_s$ are the corresponding constituent quark masses of $K_0^\ast(1430)$, and we take $\hat{m}_s = 370\ {\rm MeV}$ and $\hat{m}_q = 250\ {\rm MeV}$ as discussed in Ref.~\cite{Zhong:2022ecl}. The $\Psi _{2;K_0^*}^R(x,{{\rm{k}}_ \bot })$ is the spatial wave function, and which can be divided into the $x$-dependent part, i.e., ${\varphi _{2;K_0^*}}(x)$, dominating WF's longitudinal distribution, and the $\mathbf{k}_\perp$-dependent part arising from harmonic oscillator solution for meson in the rest frame. Then, the $\Psi _{2;K_0^\ast}^R(x,\mathbf{k}_\perp)$ can be written as:
\begin{align}
\Psi _{2;K_0^\ast}^R(x,\mathbf{k}_\perp) &= A_{2;K_0^\ast} \varphi_{2;K_0^\ast}(x) \nonumber\\
&\times \exp \Big[ - \frac{1}{8 \beta_{2;K_0^\ast}^2} \Big( \frac{\mathbf{k}_\perp^2 + \hat{m}_s^2}{x} + \frac{\mathbf{k}_\perp^2+ \hat{m}_q^2}{\bar{x}} \Big) \Big]
\label{eq:spatialWF}
\end{align}
with
\begin{align}
\varphi_{2;K_0^\ast}(x) &= (x\bar{x})^{\alpha_{2;K_0^\ast}} \Big[ C_1^{3/2}(2x - 1) + \hat{B}_{2;K_0^\ast} C_2^{3/2}(2x - 1) \Big],
\label{eq:varphiWF}
\end{align}
where ${A_{2;K_0^\ast}}$ is the normalization constant, ${\beta _{2;K_0^\ast}}$ is the harmomous parameters that dominates the WF's transverse distribution, $C_n^{3/2}(2x - 1)$ is the Gegenbauer polynomial. Considering the $SU_f(3)$ breaking effect of $K_0^\ast(1430)$, we introduce a term proportional to $C_2^{3/2}(2x-1)$. We take $\hat{B}_{2;K_0^\ast} \simeq -0.025$ in order to make the undetermined model parameters as few as possible. The value of $\hat{B}_{2;K_0^\ast}$ is taken by referring to the ratio of the second and first $\xi$-moments calculated in Sec.~\ref{Sec:IIIA}, i.e., $\langle\xi^2\rangle_{2;K_0^\ast} /\langle\xi^1 \rangle_{2;K_0^\ast}$, and whose rationality can be judged by the goodness of fit.

There is a relationship between the $K_0^\ast(1430)$ leading-twist DA and its WF,
\begin{align}
\phi _{2;K_0^\ast}(x,\mu) = \int_{|\mathbf{k}_\perp|^2 \le \mu^2} \frac{d^2\mathbf{k}_\perp}{16\pi^3} \Psi_{2;K_0^\ast}(x,\mathbf{k}_\perp).
\label{eq:DAWF}
\end{align}
Substituting the WF formula~\eqref{eq:WF} with Eqs.~\eqref{eq:spinWF}, \eqref{eq:spatialWF} and \eqref{eq:varphiWF} into~\eqref{eq:DAWF} and after integrating over the transverse momentum $\mathbf{k}_\perp$, the $K_0^\ast(1430)$ leading-twist DA reads
\begin{align}
\phi_{2;K_0^\ast}(x,\mu) &= \frac{A_{2;K_0^\ast} \beta_{2;K_0^\ast} \widetilde{m}}{4\sqrt{2} \pi^{3/2}} \sqrt{x\bar{x}} \varphi_{2;K_0^\ast}(x) \nonumber\\
&\times \exp \Big[ -\frac{\hat{m}_q^2 x + \hat{m}_s^2 \bar{x} - \widetilde{m}^2}{8\beta_{2;K_0^\ast}^2 x\bar{x}} \Big] \nonumber\\
&\times \Big\{ {\rm Erf} \Big( \sqrt{\frac{\widetilde{m}^2 + \mu^2}{8\beta_{2;K_0^\ast}^2 x\bar{x}}} \Big) - {\rm Erf} \Big( \sqrt{\frac{\widetilde{m}^2}{8\beta_{2;K_0^\ast}^2 x\bar{x}}} \Big) \Big\}
\label{eq:DA}
\end{align}
with the error function ${\rm Erf}(x) = 2\int^x_0 dt e^{-t^2}/\sqrt{\pi}$. It can be seen from the above derivation process that the scale dependence of DA $\phi_{2;K_0^\ast}(x,\mu)$ is derived from the upper limit of the transverse momentum integral in Eq.~\eqref{eq:DAWF} on the one hand (which causes the scale $\mu$ to appear in the error function explicitly), and from the scale dependence of wave function $\Psi_{2;K_0^\ast}(x,\mathbf{k}_\perp)$ on the other hand (carried by model parameters $A_{2;K_0^\ast}$, $\beta_{2;K_0^\ast}$ and $\alpha_{2;K_0^\ast}$). In the numerical calculation of determining the behavior of $\phi_{2;K_0^\ast}(x,\mu)$, the scale of DA $\phi_{2;K_0^\ast}(x,\mu)$ matches the corresponding scale of the values of $\xi$-moments $\langle\xi^n\rangle_{2;K_0^\ast}$ via the definition \eqref{xinT2DA}.

It needs to be clear that the LCHO models for WF $\Psi_{2;K_0^\ast}(x, \mathbf{k}_\perp)$ and DA $\phi_{2;K_0^\ast}(x,\mu)$ established above are the same for $K_0^\ast(1430)^+$ and $K_0^\ast(1430)^0$ due to the isospin symmetry between the $u$ and $d$ quarks. The leading-twist WF and DA of $\overline{K}_0^\ast(1430)^0$ and $K_0^\ast(1430)^-$ can be obtained by replacing $x$ with $\bar{x}$ in Eqs.~\eqref{eq:WF} and~\eqref{eq:DA}.

Now, there are three unknown model parameters such as $A_{2;K_0^\ast}$, $\beta_{2;K_0^\ast}$ and $\alpha_{2;K_0^\ast}$. In order to definitively describe the behavior of $K_0^\ast(1430)$ leading-twist DA with the LCHO model~\eqref{eq:DA}, these three parameters can be determined by fitting the $\xi$-moments with the least squares method as fitting parameters. For specific fitting procedure, one can refer to Refs.~\cite{Zhong:2021epq, Zhong:2022ecl}.

\subsection{$B_s,D_s\to K_0^\ast$ TFFs within LCSRs}

In order to uniformly express the derivation process of the LCSRs for $B_s,D_s \to K_0^\ast(1430)$ TFFs, we introduce the following vacuum-to-$K_0^\ast(1430)$ correlators
\begin{align}
\Pi_\mu(p,q) &= i\int d^4x e^{iq\cdot x} \langle K_0^\ast(p)| T\{ \bar{q}_2(x)\gamma_\mu \gamma_5 Q(x) \nonumber\\
&\times \bar{Q}(0) i\gamma_5 q_1(0) \} |0\rangle \nonumber\\
&= F(q^2,(p+q)^2)p_\mu + \widetilde{F}(p^2,(p+q)^2)q_\mu, \nonumber\\
\widetilde{\Pi}_\mu(p,q) &= i\int d^4x e^{iq\cdot x} \langle K_0^\ast(p)| T\{ \bar{q}_2(x)\sigma_{\mu\nu} \gamma_5 q^\nu Q(x) \nonumber\\
&\times \bar{Q}(0) i\gamma_5 q_1(0) \} |0\rangle \nonumber\\
&= F^T(p^2,(p+q)^2) [p_\mu q^2 - q_\mu (p\cdot q)].
\label{eq:correlatorTFFs}
\end{align}
In Eq.~\eqref{eq:correlatorTFFs}, the light quark $q_1 = s, q_2 = u$ and the heavy quark $Q = b$ is for $H_{q_1}(=B_s)\to K_0^\ast(1430)$ decay; the light quark $q_1 = s, q_2 = d$ and the heavy quark $Q = c$ is for $H_{q_1}(=D_s)\to K_0^\ast(1430)$ decay, respectively.

We first calculate the correlator~\eqref{eq:correlatorTFFs} in QCD. At $q^2\ll m_Q^2$ and $(p+q)^2 \ll m_Q^2$ with the heavy quark mass $m_Q$, the heavy quark propagating in the correlator is highly virtual and the distances are near the light-cone~\cite{Duplancic:2008ix}. Thus one can contract the heavy quark fields, and the following light-cone expansion of the heavy quark propagator,
\begin{align}
&\langle 0|Q_\alpha^i(x)\bar{Q}_\beta^j(0)|0\rangle \nonumber\\
&\quad\quad = i\int \frac{d^4k}{(2\pi)^4} e^{-ik\cdot x} \Big[ \delta^{ij} \frac{\slashed{k} + m_Q}{k^2 - m_Q^2} + \cdots \Big]_{\alpha\beta},
\label{eq:HQpropagator}
\end{align}
enters the correlator. The vacuum-to-$K_0^\ast(1430)$ matrix element can be expanded in terms of the $K_0^\ast(1430)$ light-cone DA's of growing twist. That is,
\begin{align}
&\langle K_0^\ast(p)|\bar{q}_{2\alpha}^i(x) q_{1\beta}^j(0)|0\rangle \nonumber\\
&= \frac{\delta^{ji}}{12} \bar{f}_{K_0^\ast} \int^1_0 du e^{iup\cdot x} \Big\{ \slashed{p}\phi_{2;K_0^\ast}(u) + m_{K_0^\ast} \phi_{3;K_0^\ast}^s(u) \nonumber\\
&- \frac{1}{6} m_{K_0^\ast} \sigma_{\mu\nu} p^\mu x^\nu \phi_{3;K_0^\ast}^\sigma(u) \Big\}_{\beta\alpha} + \cdots
\label{eq:K0starExpansion}
\end{align}
with the $K_0^\ast(1430)$ two-particle twist-3 DA $\phi_{3;K_0^\ast}^\sigma(u,\mu)$. In which, the Fock components of the $K_0^\ast(1430)$ with multiplicities larger than two as well as the twists higher than 3 are neglected due to that only the free propagator is retained in Eq.~\eqref{eq:HQpropagator}. This truncation is reasonable and has to be done because we almost know nothing about those components and their contributions are usually small. Then the OPE for the invariant amplitudes $F$, $\widetilde{F}$ and $F^T$ can be obtained as
\begin{align}
&F_{\rm qcd}(q^2,(p+q^2)) \nonumber\\
&\quad\quad= i\bar{f}_{K_0^\ast} m_Q \int^1_0 \frac{du}{m_Q^2 - (up+q)^2} \Big\{ \phi_{2;K_0^\ast}(u) \nonumber\\
&\quad\quad- \frac{m_{K_0^\ast}}{m_Q} u \phi^p_{3;K_0^\ast}(u) - \frac{m_{K_0^\ast}}{6m_Q} \Big[ 2 + \frac{m_Q^2 + q^2 - u^2p^2}{m_Q^2 - (up+q)^2} \Big] \nonumber\\
&\quad\quad\times \phi_{3;K_0^\ast}^\sigma(u) \Big\}, \nonumber\\
&\widetilde{F}_{\rm qcd}(q^2,(p+q)^2) \nonumber\\
&\quad\quad= i\bar{f}_{K_0^\ast} \int^1_0 \frac{du}{m_Q^2 - (up+q)^2} \Big\{ -m_{K_0^\ast} \phi_{3;K_0^\ast}^p(u) \nonumber\\
&\quad\quad- \frac{m_{K_0^\ast}}{6} \Big[ 1 - \frac{m_Q^2 - q^2 + u^2p^2}{m_Q^2 - (up+q)^2} \Big] \frac{\phi_{3;K_0^\ast}^\sigma(u)}{u} \Big\}, \nonumber\\
&F^T_{\rm qcd}(q^2,(p+q^2)) \nonumber\\
&\quad\quad= \bar{f}_{K_0^\ast} \int^1_0 \frac{du}{m_Q^2 - (up+q)^2} \Big\{ \phi_{2;K_0^\ast}(u) \nonumber\\
&\quad\quad- \frac{m_{K_0^\ast} m_Q}{3[m_Q^2 - (up+q)^2]} \phi_{3;K_0^\ast}^\sigma(u) \Big\},
\label{eq:OPE}
\end{align}
respectively.

One can also insert a complete set of hadronic states between the currents in correlator~\eqref{eq:correlatorTFFs} to obtain the hadronic representations of the invariant amplitudes. In which, the TFFs $f_{\pm,T}(q^2)$ enters the correlator~\eqref{eq:correlatorTFFs} via the hadronic matrix elements for the interpolating currents indicating the weak transition of $Q$ to $q_2$. They can be parameterized in terms of the TFFs $f_{\pm,T}(q^2)$ as
\begin{align}
&\langle K_0^\ast(p)|\bar{q}_2 \gamma_\mu \gamma_5 Q|H_{q_1}(p+q)\rangle \nonumber\\
&\quad\quad\quad\quad= -2if_+(q^2)p_\mu - i[f_+(q^2) + f_-(q^2)] q_\mu, \nonumber\\
&\langle K_0^\ast(p)|\bar{q}_2 \sigma_{\mu\nu} \gamma_5 q^\nu Q|H_{q_1}(p+q)\rangle \nonumber\\
&\quad\quad\quad\quad= [2p_\mu q^2 - 2q_\mu(p\cdot q)] \frac{-f_T(q^2)}{m_{H_{q_1}} + m_{K_0^\ast}}.
\label{eq:HME}
\end{align}
Otherwise, the vacuum-to-meson matrix element for the interpolating current representing the $H_{q_1}$ channel can be given by
\begin{align}
\langle H_{q_1}|\bar{Q} i\gamma_5 q_1 |0\rangle = \frac{m_{H_{q_1}}^2 f_{H_{q_1}}}{m_Q + m_{q_1}}
\label{eq:DecayConstant}
\end{align}
with the heavy meson mass $m_{H_{q_1}}$ and decay constant $f_{H_{q_1}}$. Then the hadronic representations of the invariant amplitudes $F$, $\widetilde{F}$ and $F^T$ can be written as
\begin{align}
&F_{\rm had}(p^2,(p+q)^2) \nonumber\\
&\quad\quad= \frac{-2i m_{H_{q_1}}^2 f_{H_{q_1}} f_+(q^2)}{(m_Q + m_{q_1}) [m_{H_{q_1}}^2 - (p+q)^2]} + \cdots, \nonumber\\
&\widetilde{F}_{\rm had}(p^2,(p+q)^2) \nonumber\\
&\quad\quad= \frac{-i m_{H_{q_1}}^2 f_{H_{q_1}} [f_+(q^2) + f_-(q^2)]}{(m_Q + m_{q_1}) [m_{H_{q_1}}^2 - (p+q)^2]} + \cdots, \nonumber\\
&F_{\rm had}^T(p^2,(p+q)^2) \nonumber\\
&\quad\quad= \frac{-2m_{H_{q_1}}^2 f_{H_{q_1}} f_T(q^2)}{(m_Q + m_{q_1}) (m_{H_{q_1}} + m_{K_0^\ast}) [m_{H_{q_1}}^2 - (p+q)^2]} \nonumber\\
&\quad\quad+ \cdots,
\label{eq:had}
\end{align}
respectively. In Eq.~\eqref{eq:had}, the ground state heavy meson contributions have been isolated, and the ellipses indicate the contributions from the excited states, the continuum states and possible subtraction terms.

Without losing generality, we take the invariant amplitude $F(q^2, (p+q)^2)$ as an example to illustrate the subsequent calculation procedure. One can write a general dispersion relation for $F(q^2, (p+q)^2)$ and further apply the Borel transformation with respect to the momentum squared $(p+q)^2$ of the heavy meson~\cite{Belyaev:1993wp},
\begin{align}
F(q^2,M^2) = \int^\infty_{t_{min}} \rho(q^2,s) e^{-s/M^2} ds
\label{eq:DispersionRelationTFFs}
\end{align}
with $t_{min} = (m_Q+m_{q_1})^2$. In which, the spectral density is given by
\begin{align}
\rho(q^2,s) &= \frac{1}{\pi} {\rm Im} F_{\rm had}(q^2,s) \nonumber\\
&= \delta(s-m_{H_{q_1}}^2) \frac{-2i m_{H_{q_1}}^2 f_{H_{q_1}} f_+(q^2)}{m_Q + m_{q_1}} \nonumber\\
&+ \frac{1}{\pi} {\rm Im} F_{\rm qcd}(q^2,s) \theta(s - s_{H_{q_1}}),
\label{eq:SpectralDensity}
\end{align}
where the contributions of the excited states and continuum states in $F_{\rm had}(q^2,s)$ have been parameterized as ${\rm Im} F_{\rm qcd}(q^2,s)/\pi$ and been delimited by the effective threshold parameter $s_{H_{q_1}}$ with the quark-hadronic duality approximation, and the possible subtractions will be got rid of due to Borel transformation in Eq.~\eqref{eq:DispersionRelationTFFs}. Substituting Eq.~\eqref{eq:SpectralDensity} into Eq.~\eqref{eq:DispersionRelationTFFs}, one can get
\begin{align}
F(q^2,M^2) &= \frac{-2i m_{H_{q_1}}^2 f_{H_{q_1}} f_+(q^2)}{m_Q + m_{q_1}} e^{-m_{H_{q_1}}^2/M^2} \nonumber\\
&+ \frac{1}{\pi} \int^\infty_{s_{H_{q_1}}} {\rm Im} F_{\rm qcd}(q^2, s) e^{-s/M^2} ds.
\label{eq:LCSRhad}
\end{align}
On the other hand, the invariant amplitude after Borel transformation can also be written as~\cite{Belyaev:1993wp}
\begin{align}
F(q^2,M^2) &= \frac{1}{\pi} \int^\infty_{t_{min}} {\rm Im} F_{\rm qcd}(q^2, s) e^{-s/M^2} ds.
\label{eq:LCSRqcd}
\end{align}
Finally, equating Eqs.~\eqref{eq:LCSRhad} with~\eqref{eq:LCSRqcd}, the LCSR of TFF $f_+(q^2)$ can be obtained as
\begin{align}
f_+(q^2) &= \frac{(m_Q+m_{q_1}) \bar{f}_{K_0^\ast} e^{m_{H_{q_1}}^2/M^2}}{2m_{H_{q_1}}^2 f_{H_{q_1}}} \nonumber\\
&\times \int^{\widetilde{u}_0}_{u_0} du e^{-(m_Q^2-\bar{u}q^2+u\bar{u}m_{K_0^\ast}^2)/(uM^2)} \nonumber\\
&\times \Big\{ -m_Q \frac{\phi_{2;K_0^\ast}(u)}{u} + m_{K_0^\ast} \phi_{3;K_0^\ast}^p(u) \nonumber\\
&+ m_{K_0^\ast} \Big[ \frac{2}{u} + \frac{4u m_Q^2 m_{K_0^\ast}^2}{(m_Q^2-q^2+u^2m_{K_0^\ast}^2)^2} \nonumber\\
&- \frac{m_Q^2+q^2-u^2m_{K_0^\ast}^2}{m_Q^2-q^2+u^2m_{K_0^\ast}^2}  \frac{d}{du} \Big] \frac{\phi_{3;K_0^\ast}^\sigma(u)}{6} \Big\}. \label{eq:LCSRTFFs1}
\end{align}
Similarly,
\begin{align}
f_+(q^2) &+ f_-(q^2) = \frac{(m_Q+m_{q_1}) \bar{f}_{K_0^\ast} m_{K_0^\ast} e^{m_{H_{q_1}}^2/M^2}}{m_{H_{q_1}}^2 f_{H_{q_1}}} \nonumber\\
&\times \int^{\widetilde{u}_0}_{u_0} du e^{-(m_Q^2-\bar{u}q^2+u\bar{u}m_{K_0^\ast}^2)/(uM^2)} \nonumber\\
&\times \Big[ \frac{\phi_{3;K_0^\ast}^p(u)}{u} + \frac{1}{u} \frac{d}{du} \frac{\phi_{3;K_0^\ast}^\sigma(u)}{6} \Big], \label{eq:LCSRTFFs2}\\
f_T(q^2) &= \frac{(m_{H_{q_1}} + m_{K_0^\ast}) (m_Q+m_{q_1}) \bar{f}_{K_0^\ast} e^{m_{H_{q_1}}^2/M^2}}{m_{H_{q_1}}^2 f_{H_{q_1}}} \nonumber\\
&\times \int^{\widetilde{u}_0}_{u_0} du e^{-(m_Q^2-\bar{u}q^2+u\bar{u}m_{K_0^\ast}^2)/(uM^2)} \nonumber\\
&\times \Big\{ -\frac{\phi_{2;K_0^\ast}(u)}{2u} + \frac{m_Q m_{K_0^\ast}}{m_Q^2-q^2+u^2m_{K_0^\ast}^2} \nonumber\\
&\times \Big[ \frac{2um_{K_0^\ast}^2}{m_Q^2-q^2+u^2m_{K_0^\ast}^2} - \frac{d}{du} \Big] \frac{\phi_{3;K_0^\ast}^\sigma(u)}{6} \Big\}
\label{eq:LCSRTFFs3}
\end{align}
with
\begin{align}
u_0 &= \Big[ \sqrt{(q^2 - s_{H_{q_1}} + m_{K_0^\ast}^2)^2 + 4m_{K_0^\ast}^2 (m_Q^2 - q^2)} \nonumber\\
&+ q^2 - s_{H_{q_1}} + m_{K_0^\ast}^2 \Big]/(2m_{K_0^\ast}^2), \\
\widetilde{u}_0 &= \Big[ \sqrt{(q^2 - t_{min} + m_{K_0^\ast}^2)^2 + 4m_{K_0^\ast}^2 (m_Q^2 - q^2)} \nonumber\\
&+ q^2 - t_{min} + m_{K_0^\ast}^2 \Big]/(2m_{K_0^\ast}^2).
\end{align}
In particular, the upper limit of the integral of variable $u$ in LCSRs~\eqref{eq:LCSRTFFs1},~\eqref{eq:LCSRTFFs2} and~\eqref{eq:LCSRTFFs3} is $\widetilde{u}_0$ instead of 1, because the lower limit of the integral variable $s$ in the dispersion relationship~\eqref{eq:DispersionRelationTFFs}, $t_{min}$, is larger than, but not equal to, $m_Q^2$.

\section{NUMERICAL ANALYSIS}

\subsection{$\xi$-moments and behavior of $\phi_{2;K_0^\ast}(u,\mu)$}\label{Sec:IIIA}

Now we can calculate the values of the $\xi$-moments of $K_0^\ast(1430)$ leading-twist DA. In calculation, we take the mass of $K_0^\ast(1430)$ as $m_{K_0^\ast} = 1.425_{-0.050}^{+0.050}$ $\rm{GeV}$, the $u$ and $s$ current quark mass are adopted as $m_u = 2.16_{-0.26}^{+0.49}$ $\rm{MeV}$ and $m_s = 93_{-5}^{+11}$ $\rm{MeV}$ at scale $\mu = 2$ $\rm{GeV}$, respectively~\cite{Workman:2022ynf}. For the decay constant of $K_0^\ast(1430)$, we take the value from QCD sum rule calculation, that is, $f_{K_0^\ast} = 427_{-85}^{+85}$ $\rm{MeV}$ at $\mu \sim 1\ {\rm GeV}$~\cite{Du:2004ki}. The values of the scale dependent vacuum condensates are: $\langle\bar{u}u\rangle = (-2.417_{-0.114}^{+0.227}) \times 10^{-2}~\rm{GeV}^3$, $\langle\bar{s}s\rangle = \kappa \langle\bar{u}u\rangle$ with $\kappa = 0.74\pm 0.03$, $\langle g_s\bar{u}\sigma TGu \rangle = (-1.934^{+0.188}_{-0.103}) \times 10^{-2}~\rm{GeV}^5$, $\langle g_s\bar{s}\sigma TGs \rangle = \kappa \langle g_s\bar{u}\sigma TGu \rangle$, $\langle g_s\bar{u}u\rangle^2 = (2.082_{-0.679}^{+0.734}) \times 10^{-3}~{\rm GeV}^6$, $\langle g_s\bar{s}s\rangle^2 = \kappa^2 \langle g_s\bar{u}u\rangle^2$ and $\langle g_s^2\bar{u}u\rangle^2 = (7.420_{-2.483}^{+2.614}) \times 10^{-3}~{\rm GeV}^6$ at $\mu = 2~{\rm GeV}$; the values of the scale independent gluon condensates are: $\langle\alpha_s G^2\rangle = 0.038_{-0.011}^{+0.011} ~{\rm GeV}^4$ and $\langle g_s^3fG^3\rangle \simeq 0.045 ~{\rm GeV}^6$~\cite{Zhong:2021epq, Zhong:2014jla, Colangelo:2000dp, Narison:2014ska, Narison:2014wqa}. In addition, we take the scale $\mu = M$ in sum rules~\eqref{SRxin} as usual, and the continuum threshold $s_{K_0^\ast} \simeq 7.8~{\rm GeV}^2$ by requiring that there is a reasonable Borel window to normalize $\langle\xi^0_p\rangle_{3;K_0^\ast}$ in Eq.~\eqref{SRxi0xi0} as suggested in Ref.~\cite{Zhong:2021epq}. The scale evolutions of $\xi$-moments are realized by that of the Gegenbauer moments via the relationship between both of them. For the scale evolutions of Gegenbauer moments, one can refer to Ref.~\cite{Cheng:2005nb}.

Substituting the above inputs into Eq.~\eqref{SRxin}, the $\xi$-moments $\langle\xi^n\rangle_{2;K_0^\ast}$, continuum state contributions and dimension-six term contributions versus Borel parameter $M^2$ can be obtained. We will evaluate the values of the first ten $\xi$-moments in this work. In order to estimate these values, one should determine the appropriate Borel windows. We require the continuum state contributions are not more than $30\%$, $35\%$, $40\%$, $45\%$, $50\%$ for odd $\xi$-moments $\langle\xi^n\rangle_{2;K_0^\ast} (n=1,3,5,7,9)$, respectively, to get the upper limits of the corresponding Borel windows. On the other hand, the dimension-six term contributions for those five odd $\xi$-moments are far less than $5\%$ in a very wide Borel parameter region, then the basic criteria that require dimension-six term contributions to be as small as possible are automatically satisfied. Reasonably, we directly fix the lengths of the corresponding Borel windows as $1~{\rm GeV}^2$. Due to that the even $\xi$-moments are very close to zero, one cannot determine the Borel windows by limiting the continuum state contributions and dimension-six term contributions, so we determine their Borel windows by examining the stability of even $\xi$-moments changes with Borel parameters.

\begin{figure}[htb]
\begin{center}
\includegraphics[width=0.43\textwidth]{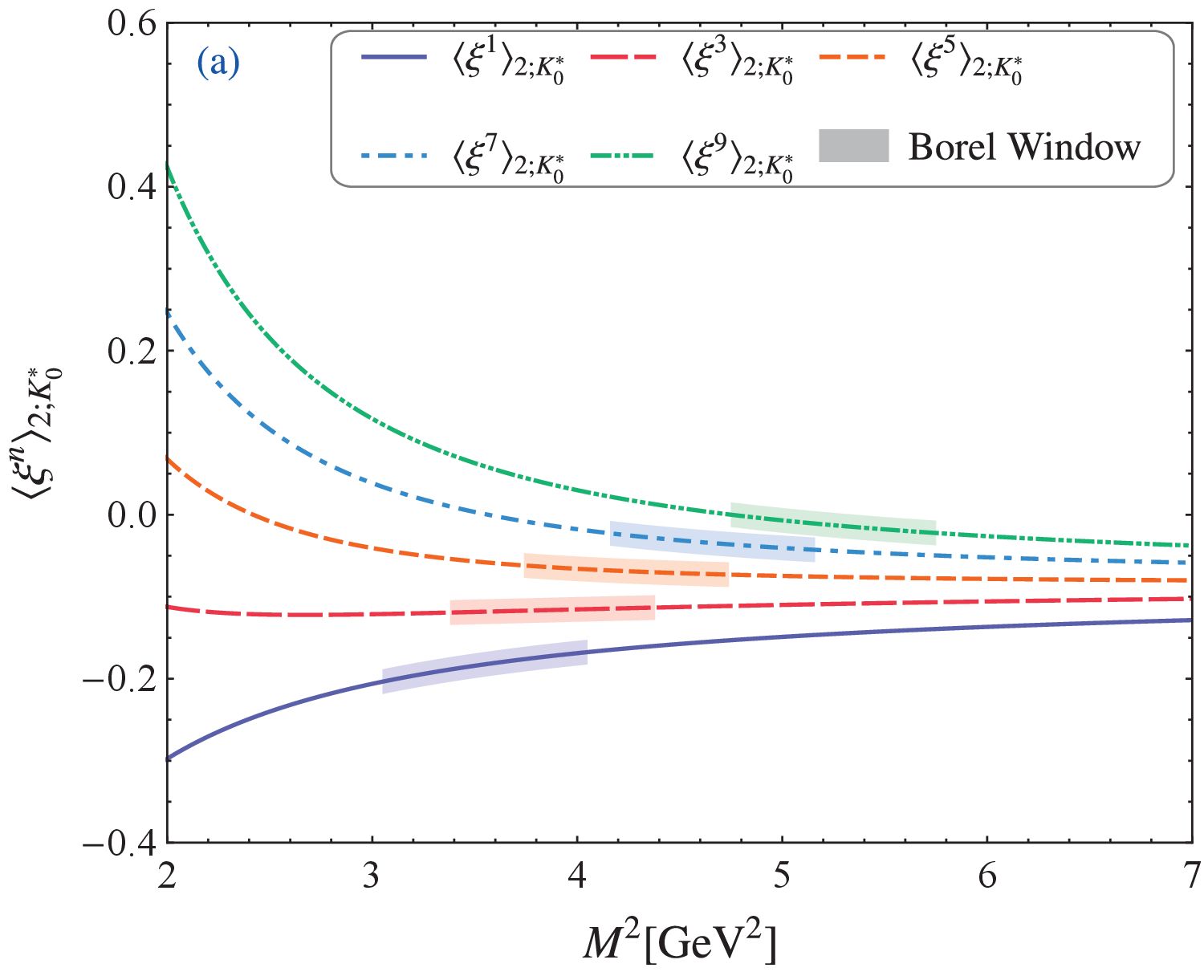}
\includegraphics[width=0.43\textwidth]{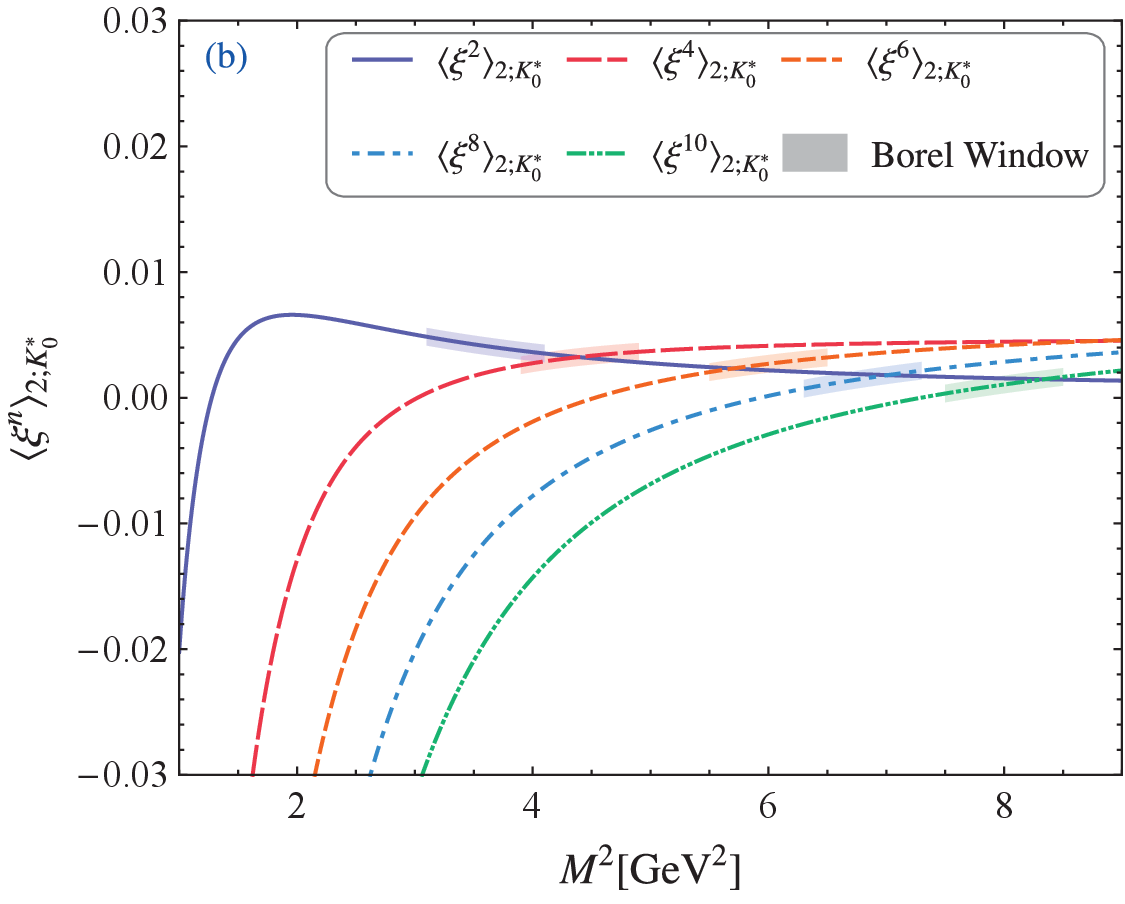}
\end{center}
\caption{$K_0^\ast(1430)$ leading-twist DA $\xi$-moments $\langle\xi^n\rangle_{2;K_0^\ast} (n = 1,2,\cdots,10)$ versus the Borel parameter $M^2$, where the shaded bands indicate the Borel windows, and all input parameters are set to be their central values.}
\label{fig:xin}
\end{figure}
\begin{table}[t]
\footnotesize
\begin{center}
\caption{Our predictions for the first ten $\xi$-moments $\langle\xi^n\rangle_{2;K_0^\ast} (n = 1,2,\cdots,10)$ of $K_0^\ast(1430)$ leading-twist DA with the scale $\mu = 1, 1.4$ and $3~{\rm GeV}$.}
\label{table:xin1}
\begin{tabular}{l c c c}
\hline\hline
$\langle\xi^n\rangle_{2;K_0^\ast}$~~~~~ &~~~~~~$1$~$\rm{GeV}$~~~~~~ &~~~~~~$1.4$~$\rm{GeV}$~~~~~~ &~~~~~~$3$~$\rm{GeV}$~~~~~ \\
\hline
$\langle \xi^1 \rangle _{2;K_0^\ast}$    &$-0.261_{-0.071}^{+0.056}$    &$-0.211_{-0.057}^{+0.045}$    &$-0.156_{-0.043}^{+0.033}$    \\
$\langle \xi^2 \rangle _{2;K_0^\ast}$    &$~~0.0065_{-0.0057}^{+0.0046}$   &$~~0.0050_{-0.0043}^{+0.0035}$  &$~~0.0034_{-0.0030}^{+0.0024}$  \\
$\langle \xi^3 \rangle _{2;K_0^\ast}$    &$-0.177_{-0.045}^{+0.034}$    &$-0.138_{-0.035}^{+0.026}$    &$-0.098_{-0.025}^{+0.019}$    \\
$\langle \xi^4 \rangle _{2;K_0^\ast}$    &$~~0.0052_{-0.0037}^{+0.0031}$   &$~~0.0039_{-0.0028}^{+0.0024}$ &$~~0.0026_{-0.0019}^{+0.0016}$   \\
$\langle \xi^5 \rangle _{2;K_0^\ast}$    &$-0.103_{-0.028}^{+0.022}$    &$-0.081_{-0.022}^{+0.017}$    &$-0.058_{-0.016}^{+0.012}$    \\
$\langle \xi^6 \rangle _{2;K_0^\ast}$    &$~~0.0044_{-0.0027}^{+0.0024}$   &$~~0.0033_{-0.0020}^{+0.0018}$ &$~~0.0022_{-0.0014}^{+0.0012}$   \\
$\langle \xi^7 \rangle _{2;K_0^\ast}$    &$-0.045_{-0.022}^{+0.021}$    &$-0.038_{-0.017}^{+0.016}$    &$-0.030_{-0.012}^{+0.011}$    \\
$\langle \xi^8 \rangle _{2;K_0^\ast}$    &$~~0.0025_{-0.0026}^{+0.0024}$   &$~~0.0020_{-0.0019}^{+0.0017}$ &$~~0.0014_{-0.0013}^{+0.0011}$   \\
$\langle \xi^9 \rangle _{2;K_0^\ast}$    &$-0.018_{-0.023}^{+0.023}$    &$-0.018_{-0.017}^{+0.016}$    &$-0.016_{-0.011}^{+0.010}$    \\
$\langle \xi^{10} \rangle _{2;K_0^\ast}$ &$~~0.0018_{-0.0024}^{+0.0021}$   &$~~0.0014_{-0.0017}^{+0.0015}$ &$~~0.0010_{-0.0011}^{+0.0010}$   \\
\hline\hline
\end{tabular}
\end{center}
\end{table}
\begin{table*}[t]
\centering
\caption{Our predictions for the first three $\xi$-moments and Gegenbauer moments of $K_0^\ast(1430)$ leading-twist DA at scale $\mu = 1~{\rm GeV}$, compared to other theoretical predictions.}\label{table:xin_value}
\begin{tabular}{l c c c c c c}
\hline\hline
~& ~~~~~~$\langle\xi^1\rangle_{2;K_0^\ast}$~~~~~~ & ~~~~~~$\langle\xi^2\rangle_{2;K_0^\ast}$~~~~~~ & ~~~~~~$\langle\xi^3\rangle_{2;K_0^\ast}$~~~~~~ & ~~~~~~$a_1^{2;K_0^\ast}$~~~~~~ & ~~~~~~$a_2^{2;K_0^\ast}$~~~~~~ & ~~~~~~$a_3^{2;K_0^\ast}$~~~~~~  \\
\hline
This Work               & $-0.261_{-0.071}^{+0.056}$ & $~~0.0065_{-0.0057}^{+0.0046}$ & $-0.177_{-0.045}^{+0.034}$ & $-0.435_{-0.118}^{+0.093}$ & $~~0.019_{-0.017}^{+0.014}$ & $-0.342_{-0.076}^{+0.051}$ \\
QCD SR~\cite{Cheng:2005nb}       & $-0.35^{+0.08}_{-0.08}$ & $-$ & $-0.23^{+0.06}_{-0.06}$ & $-0.57^{+0.13}_{-0.13}$ & $-$ & $-0.42^{+0.22}_{-0.22}$ \\
LF Holographic~\cite{Chen:2021oul}  & $-0.078^{+0.018}_{-0.018}$ & $-0.010^{+0.001}_{-0.001}$ & $\sim -0.034$ & $-0.130^{+0.030}_{-0.030}$ & $-0.030^{+0.002}_{-0.002}$ & $-0.005^{+0.001}_{-0.001}$  \\
\hline\hline
\end{tabular}
\end{table*}
The first ten $\xi$-moments of $K_0^\ast(1430)$ leading-twist DA versus the Borel parameter and the corresponding Borel windows are shown in Fig.~\ref{fig:xin}, where all inputs are taken to be their central values. By taking all error sources into account, the values of the first ten $\xi$-moments at the scale $\mu = 1~{\rm GeV}$, $1.4~{\rm GeV}$ and $3~{\rm GeV}$ can be obtained, and which are exhibited in Table~\ref{table:xin1}. In Table~\ref{table:xin1}, the values at $\mu = 1.4~{\rm GeV}$ and $3~{\rm GeV}$ will be used for subsequent calculations of the $D_s\to K_0^\ast(1430)$ and $B_s\to K_0^\ast(1430)$ TFFs, respectively. Our values for the first three $\xi$-moments and Gegenbauer moments at scale $\mu = 1~{\rm GeV}$ are also exhibited in Table~\ref{table:xin_value}, the other theoretical predictions such as by traditional QCDSRs~\cite{Cheng:2005nb} and LF approach~\cite{Chen:2021oul} are displayed for comparison. By only adopting the central values, we can further get
\begin{align}
\frac{\langle\xi^2\rangle_{2;K_0^\ast}}{\langle\xi^1\rangle_{2;K_0^\ast}} = -0.025, \quad \frac{a_2^{2;K_0^\ast}}{a_1^{2;K_0^\ast}} = -0.044,
\end{align}
at $\mu = 1~{\rm GeV}$.

\begin{table}[t]
\footnotesize
\begin{center}
\caption{The fitting parameters and Goodness of fit of the LCHO model, when $\hat{m}_{q}=0.25~{\rm{GeV}}$, $\hat m_s$ takes different values.}
\label{table:ms Parameters}
\begin{tabular}{l c c c c c}
\hline\hline
$\hat{m}_{s} [\rm GeV]$~&$A_{2;K_0^\ast} [{\rm GeV}^{-1}]$~ &$\alpha_{2;K_0^\ast}$~ &$\beta_{2;K_0^\ast} [{\rm GeV}]$~ &~$\chi_{\rm min}^2$~&~$P_{\chi_{\rm min}^2}$~ \\
\hline
~$0.37$    &$-177$    &$-0.079$    &$0.975$    &$4.42008$    &$0.730318$    \\
~$0.36$    &$-175$    &$-0.087$    &$0.971$    &$4.25756$    &$0.749679$    \\
~$0.35$    &$-162$    &$-0.106$    &$0.991$    &$4.15859$    &$0.761341$    \\
~$0.34$    &$-179$    &$-0.082$    &$0.968$    &$3.9511$     &$0.785392$    \\
~$0.33$    &$-191$    &$-0.065$    &$0.957$    &$3.78239$    &$0.804473$    \\
~$0.32$    &$-202$    &$-0.047$    &$0.954$    &$3.62995$    &$0.821275$    \\
~$0.31$    &$-216$    &$-0.025$    &$0.951$    &$3.48259$    &$0.837066$    \\
~$0.30$    &$-227$    &$-0.009$    &$0.95$     &$3.34936$    &$0.850914$    \\
~$0.29$    &$-239$    &$0.008$     &$0.952$    &$3.24209$    &$0.86174$     \\
~$0.28$    &$-265$    &$0.086$     &$1.035$    &$3.14826$    &$0.870957$    \\
~$0.27$    &$-266$    &$0.083$     &$1.042$    &$3.14235$    &$0.871528$    \\
~$0.26$    &$-256$    &$0.062$     &$1.046$    &$3.30848$    &$0.855075$    \\
~$0.25$    &$-233$    &$0.017$     &$1.045$    &$3.71975$    &$0.811432$    \\
\hline\hline
\end{tabular}
\end{center}
\end{table}

\begin{table}[t]
\footnotesize
\begin{center}
\caption{Fitted LCHO model parameters and the corresponding goodness of fit for the $K_0^\ast(1430)$ leading-twist DA with the scale $\mu = 1, 1.4$ and $3~{\rm GeV}$, respectively.}
\label{table:DA Parameters}
\begin{tabular}{l c c c c c}
\hline\hline
$\mu [\rm GeV]$~&$A_{2;K_0^\ast} [{\rm GeV}^{-1}]$~ &$\alpha_{2;K_0^\ast}$~ &$\beta_{2;K_0^\ast} [{\rm GeV}]$~ &~$\chi_{\rm min}^2$~&~$P_{\chi_{\rm min}^2}$~ \\
\hline
~$1$    &$-266$    &$0.083$    &$1.042$    &$3.14235$    &$0.871528$     \\
~$1.4$    &$-147$    &$0.011$    &$1.091$    &$2.545$    &$0.923671$     \\
~$3$    &$-67$    &$-0.148$    &$1.088$    &$1.98132$    &$0.960867$     \\
\hline\hline
\end{tabular}
\end{center}
\end{table}
\begin{figure}[htb]
\begin{center}
\includegraphics[width=0.45\textwidth]{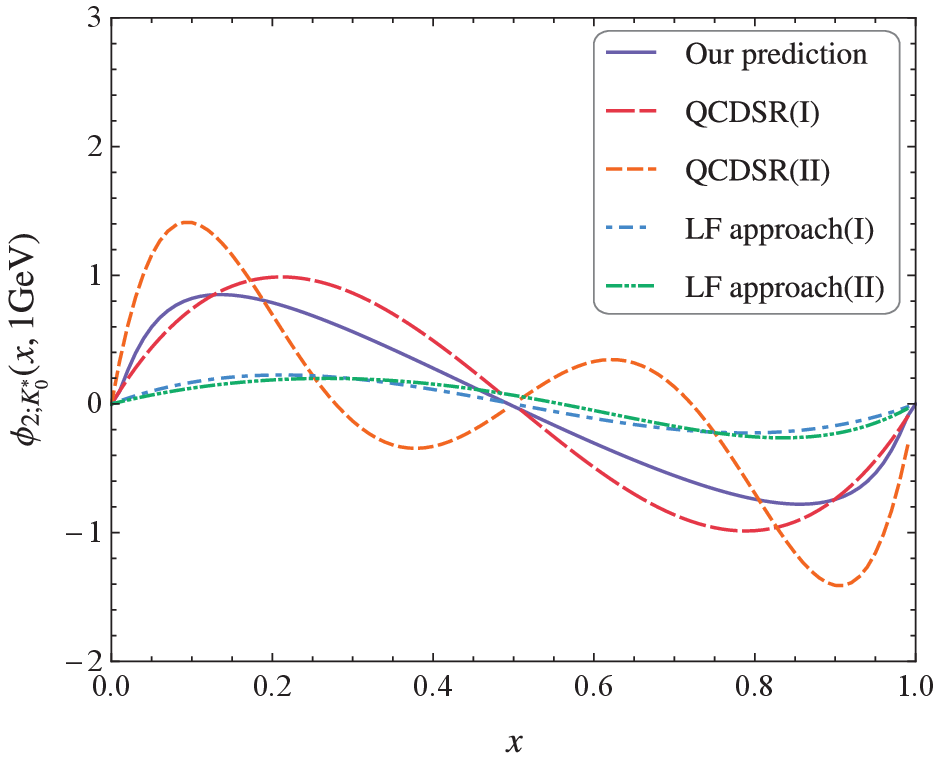}
\end{center}
\caption{Our prediction for the behavior of $K_0^\ast(1430)$ leading-twist DA at scale $\mu=1~{\rm GeV}$. We also present other predictions from the traditional QCDSRs~\cite{Cheng:2005nb} and LF approach~\cite{Chen:2021oul} for comparison.}
\label{fig:DA}
\end{figure}
Then, we can determine the behavior of $K_0^\ast(1430)$ leading-twist DA by fitting the values exhibited in Table~\ref{table:xin_value} with the LCHO model shown in Eq.~\eqref{eq:DA} via the least squares method. Following the fitting procedure introduced in detailed in Refs.~\cite{Zhong:2021epq, Zhong:2022ecl}, the model parameters and the corresponding goodness of fit are determined. However, the goodness of fit is not ideal. In order to obtain better fitting results, we fix $\hat{m}_q = 250~{\rm MeV}$ and change $\hat{m}_s$. The fitting results corresponding different $\hat{m}_s$ are exhibited in Table~\ref{table:ms Parameters}, and one can find that, when $\hat{m}_s = 270~{\rm MeV}$, the goodness of fit is best. Then we take the constituent quark mass $\hat{m}_q = 250~{\rm MeV}$ and $\hat{m}_s = 270~{\rm MeV}$ in subsequent calculations, respectively. Table~\ref{table:DA Parameters} displays the fitted LCHO model parameters and the corresponding goodness of fit with the scale $\mu = 1, 1.4$ and $3~{\rm GeV}$, respectively. The curve of our LCHO model for $K_0^\ast(1430)$ leading-twist DA at $\mu = 1~{\rm GeV}$ are shown in Fig.~\ref{fig:DA}. As a comparison, the curves from the traditional QCDSRs~\cite{Cheng:2005nb} and LF approach~\cite{Chen:2021oul} are also shown in this figure. In Refs~\cite{Cheng:2005nb, Chen:2021oul}, the TF model for $K_0^\ast(1430)$ leading-twist DA is adopted, which is,
\begin{align}
\phi_{2;K_0^\ast}^{{\rm TF},\mathcal{N} }(x,\mu) = 6x(1-x) \sum_{n=1}^\mathcal{N} a_n^{K_0^\ast} C_n^{3/2}(2x-1).
\label{eq:DATF}
\end{align}
Here, in order to be consistent with our matrix element definition of DA, i.e., Eq.~\eqref{T2DA}, so as to facilitate comparison, we have got rid off the decay constant $\bar{f}_{K_0^\ast}$ in ``$\Phi_{K_0^\ast}(x)$'' in Ref.~\cite{Cheng:2005nb}, and the constant $\bar{\mu}_{K_0^\ast}$ in ``$\Phi_{K_0^\ast}(x)$'' in Ref.~\cite{Chen:2021oul}, respectively. In Fig.~\ref{fig:DA}, the TF models corresponding to the truncations up to $\mathcal{N} = 1$ (I) and $3$ (II), respectively, are shown. From Fig.~\ref{fig:DA}, one can also find that, our DA $\phi_{2;K_0^\ast}(x,1~{\rm GeV})$ takes the maximum value at $x = 0.14$ ($\phi_{2;K_0^\ast}(x = 0.14,1~{\rm GeV}) = 0.850$), the minimum value at $x = 0.86$ ($\phi_{2;K_0^\ast}(x = 0.86,1~{\rm GeV}) = -0.779$), while the zero point is taken at $x \simeq 0.495$, respectively. This case indicates that $SU_f(3)$ breaking effect is insignificant for $K_0^\ast(1430)$ leading-twist DA.

\subsection{TFFs and branching fractions for the semileptonic $B_s,D_s \to K_{0}^{\ast}(1430)$ decays} \label{Sec:IIIB}

In order to calculate the TFFs of the semileptonic $B_s,D_s \to K_{0}^{\ast}(1430)$ decays with Eqs.~\eqref{eq:LCSRTFFs1},~\eqref{eq:LCSRTFFs2} and \eqref{eq:LCSRTFFs3}, we adopt the heavy meson masses $m_{\overline{B}_s^0} = 5366.92(10)~{\rm MeV}$ and $m_{D_s^+} = 1968.35(7)~{\rm MeV}$, decay constants $f_{B_s} = 266(19)~{\rm MeV}$ and $f_{D_s} = 256.0(42)~{\rm MeV}$, the $\overline{MS}$ heavy $b$ and $c$ quark masses $\bar{m}_b(\bar{m}_b) = 4.18_{-0.03}^{+0.04}~{\rm GeV}$ and $\bar{m}_c(\bar{m}_c) = 1.27\pm 0.02~{\rm GeV}$~\cite{Workman:2022ynf}, respectively. For the effective threshold parameters, we take $s_{B_s} = 38\pm 1~{\rm GeV}$, $s_{D_s} \simeq 6.5\pm 0.5~{\rm GeV}$. The Borel windows are taken as $13~{\rm GeV}^2 \leq M^2 \leq 17~{\rm GeV}^2$. The $K_0^\ast(1430)$ twist-3 DA $\phi_{3;K_0^\ast}^s(x,\mu)$ and $\phi_{3;K_0^\ast}^\sigma(x,\mu)$ adopted in this work are the TF model obtained in Ref.~\cite{Han:2013zg} with QCDSRs in the framework of BFT, which read,
\begin{align}
\phi_{3;K_0^\ast}^p(x,\mu) &= 1 + a_{1,p}^{3;K_0^\ast}(\mu) C_1^{1/2}(2x-1) \nonumber\\
&+ a_{2,p}^{3;K_0^\ast}(\mu) C_2^{1/2}(2x-1), \nonumber\\
\phi_{3;K_0^\ast}^\sigma(x,\mu) &= 6x\bar{x} [1 + a_{1,\sigma}^{3;K_0^\ast}(\mu) C_1^{3/2}(2x-1) \nonumber\\
&+ a_{2,\sigma}^{3;K_0^\ast}(\mu) C_2^{3/2}(2x-1)],
\end{align}
with $a_{1,p}^{3;K_0^\ast}(\mu) = 0.0126\pm 0.0017$, $a_{2,p}^{3;K_0^\ast}(\mu) = 0.187\pm 0.024$, $a_{1,\sigma}^{3;K_0^\ast}(\mu) = 0.0342\pm 0.0126$ and $a_{2,\sigma}^{3;K_0^\ast}(\mu) = 0.0235\pm 0.0045$ at $\mu = 1~{\rm GeV}$.

\begin{table}[t]
\footnotesize
\begin{center}
\caption{TFFs of the semileptonic $B_s\to K_0^\ast(1430)$ decays at large recoil point.} \label{table:TFFsB}
\begin{tabular}{l l l l}
\hline\hline
~Method~~~~~~ & $f_+^{B_s\to K_0^\ast}(0)$~~~~~~ & $f_-^{B_s\to K_0^\ast}(0)$~~~~~~ & $f_T^{B_s\to K_0^\ast}(0)$~ \\
\hline
~This work~  & $0.39_{-0.08}^{+0.08}$  & $-0.24_{-0.05}^{+0.05}$  & $0.43_{-0.09}^{+0.09}$  \\
~RQM~\cite{Faustov:2013ima}~  & $0.27^{+0.03}_{-0.03}$  & $-0.62^{+0.06}_{-0.06}$    \\
~pQCD~\cite{Li:2008tk}~  & $0.56^{+0.16}_{-0.13}$  & $-$  & $0.72^{+0.22}_{-0.17}$  \\
~pQCD~\cite{Zhang:2010af}~  & $0.56^{+0.07}_{-0.09}$  & $-$  & $-$   \\
~pQCD~\cite{Chen:2021oul}~  & $0.28^{+0.02}_{-0.02}$  & $-$  & $-$    \\
~QCDSR~\cite{Yang:2005bv}~  & $0.24^{+0.10}_{-0.10}$  & $-$  & $-$   \\
~QCDSR~\cite{Ghahramany:2009zz}~  & $0.25^{+0.05}_{-0.05}$  & $-0.17^{+0.04}_{-0.04}$  & $0.21^{+0.04}_{-0.04}$   \\
~LCSR~\cite{Khosravi:2022fzo}(I)~  & $0.28^{+0.11}_{-0.09}$  & $-0.23^{+0.09}_{-0.19}$  & $0.32^{+0.13}_{-0.11}$  \\
~LCSR~\cite{Khosravi:2022fzo}(II)~  & $0.41^{+0.14}_{-0.12}$  & $-0.37^{+0.11}_{-0.14}$  & $0.50^{+0.18}_{-0.14}$ \\
~LCSR~\cite{Khosravi:2022fzo}(III)~  & $0.51^{+0.16}_{-0.12}$  & $-0.51^{+0.12}_{-0.16}$  & $0.64^{+0.22}_{-0.15}$  \\
~LCSR~\cite{Wang:2008da}~  & $0.42^{+0.13}_{-0.07}$  & $-0.35$  & $0.52^{+0.18}_{-0.08}$   \\
~LCSR~\cite{Wang:2014upa, Wang:2014vra}~  & $0.458^{+0.062}_{-0.062}$  & $-0.240^{+0.058}_{-0.058}$  & $0.575^{+0.098}_{-0.098}$ \\
~LCSR~\cite{Sun:2010nv}~  & $0.44$  & $-0.44$  & $-$   \\
~LCSR~\cite{Han:2013zg}~  & $0.39^{+0.04}_{-0.04}$  & $-0.25^{+0.05}_{-0.05}$  & $0.41^{+0.04}_{-0.04}$   \\
\hline\hline
\end{tabular}
\end{center}
\end{table}
\begin{table}[t]
\footnotesize
\begin{center}
\caption{TFFs of the semileptonic $D_s\to K_0^\ast(1430)$ decays at large recoil point.} \label{table:TFFsD}
\begin{tabular}{l l l l }
\hline\hline
~Method~~~~~ & $f_+^{D_s\to K_0^\ast}(0)$~~~~~ & $f_-^{D_s\to K_0^\ast}(0)$~~~~~ & $f_T^{D_s\to K_0^\ast}(0)$~ \\
\hline
~This work~  & $0.65_{-0.13}^{+0.13}$  & $0.22_{-0.10}^{+0.10}$  & $0.58_{-0.13}^{+0.13}$   \\
~QCDSR~\cite{Yang:2005bv}~  & $0.51^{+0.20}_{-0.20}$  & $-$  & $-$  \\
~Fit to Data~\cite{Cheng:2002ai}~  & $1.02^{+0.07}_{-0.07}$  & $-$   \\
\hline\hline
\end{tabular}
\end{center}
\end{table}
With the above inputs and the $K_0^\ast(1430)$ leading-twist DA determined in previous subsection, the dependence of the TFFs for semileptonic $B_s, D_s \to K_0^\ast(1430)$ decays on $q^2$ in allowable $q^2$ regions of LCSRs can be obtained. In particular, the values of those TFFs at the large recoil point are exhibited in Table~\ref{table:TFFsB} and~\ref{table:TFFsD}. As a comparison, the corresponding values estimated with RQM~\cite{Faustov:2013ima}, pQCD~\cite{Li:2008tk, Zhang:2010af, Chen:2021oul}, three-point QCDSR~\cite{Yang:2005bv, Ghahramany:2009zz}, LCSR~\cite{Khosravi:2022fzo, Wang:2008da, Wang:2014upa, Wang:2014vra, Sun:2010nv, Han:2013zg}, etc., are also shown in Table~\ref{table:TFFsB} and~\ref{table:TFFsD}.

\begin{table*}[t]
\footnotesize
\begin{center}
\caption{Extrapolation parameters for the semileptonic $B_s,D_s\to K_0^\ast(1430)$ decays.} \label{table:TFFsExt}
\begin{tabular}{l c l l l l l l}
\hline\hline
~ & ~~~~~~~~~ & $f_+^{B_s\to K_0^\ast}(q^2)$~~~~~~~~ & $f_-^{B_s\to K_0^\ast}(q^2)$~~~~~~~~ & $f_T^{B_s\to K_0^\ast}(q^2)$~~~~~~~~ & $f_+^{D_s\to K_0^\ast}(q^2)$~~~~~~~~ & $f_-^{D_s\to K_0^\ast}(q^2)$~~~~~~~~ & $f_T^{D_s\to K_0^\ast}(q^2)$ \\
\hline
upper limit& $f_i(0)$    & $0.47$  & $-0.19$  & $0.53$ & $0.78$  & $0.31$  & $0.71$  \\
~             & $a_i$     & $1.17$  & $0.83$  & $1.04$ &  $0.91$  & $5.12$  & $0.10$  \\
~             & $b_i$     & $0.38$  & $1.58$  & $0.45$ & $0.11$  & $11.43$  & $1.12$  \\
\hline
central value& $f_i(0)$  & $0.39$  & $-0.24$  & $0.43$ & $0.65$  & $0.22$  & $0.58$  \\
~             & $a_i$     & $1.17$  & $0.83$  & $1.04$ & $0.91$  & $5.85$  & $-0.02$  \\
~             & $b_i$     & $0.38$  & $1.38$  & $0.47$ & $0.14$  & $16.12$  & $1.70$  \\
\hline
lower limit  & $f_i(0)$  & $0.31$  & $-0.29$  & $0.34$ & $0.52$  & $0.12$  & $0.45$  \\
~             & $a_i$     & $1.17$  & $0.83$  & $1.03$ & $0.91$  & $7.62$  & $-0.19$  \\
~             & $b_i$     & $0.39$  & $1.28$  & $0.49$ & $0.18$  & $29.67$  & $2.56$  \\
\hline\hline
\end{tabular}
\end{center}
\end{table*}
\begin{figure*}[htb]
\begin{center}
~\includegraphics[width=0.32\textwidth]{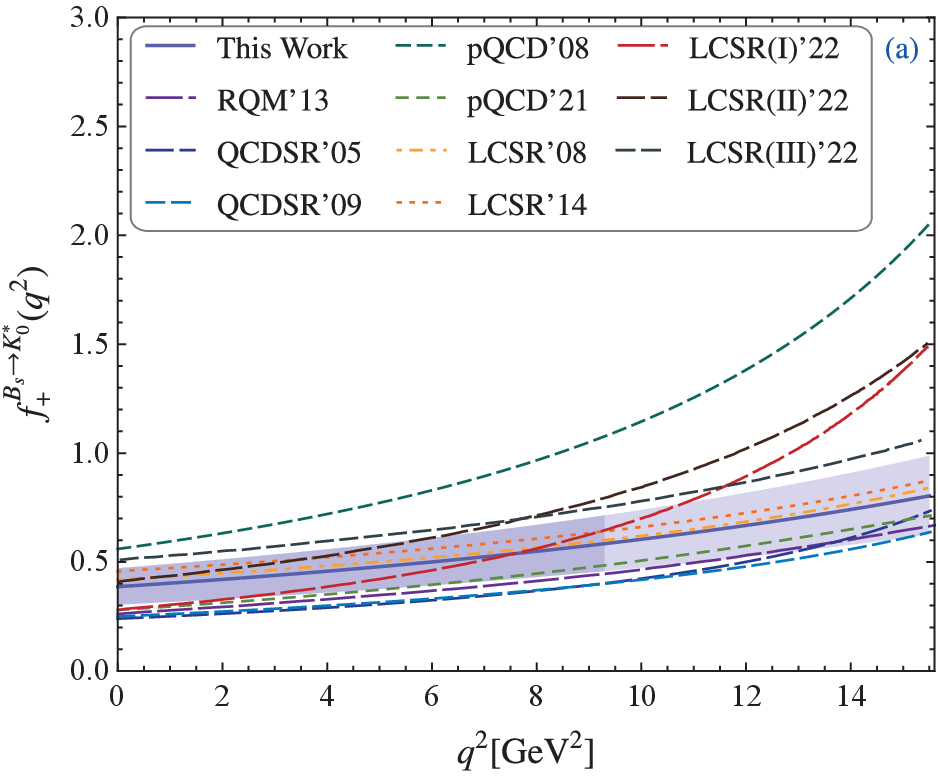}~\includegraphics[width=0.32\textwidth]{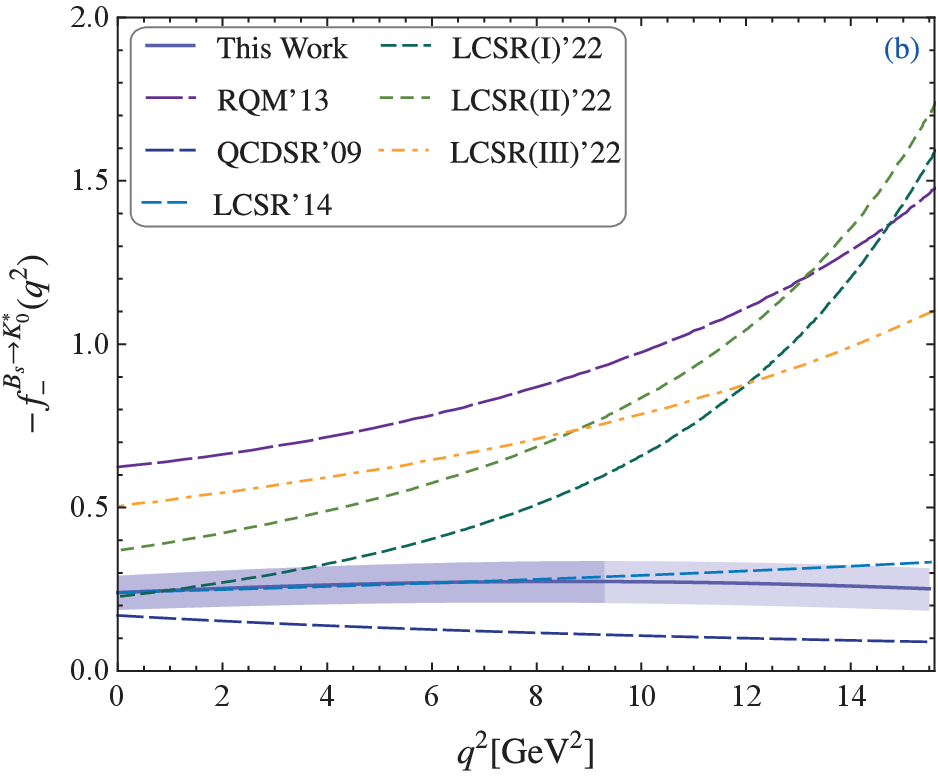}
~\includegraphics[width=0.32\textwidth]{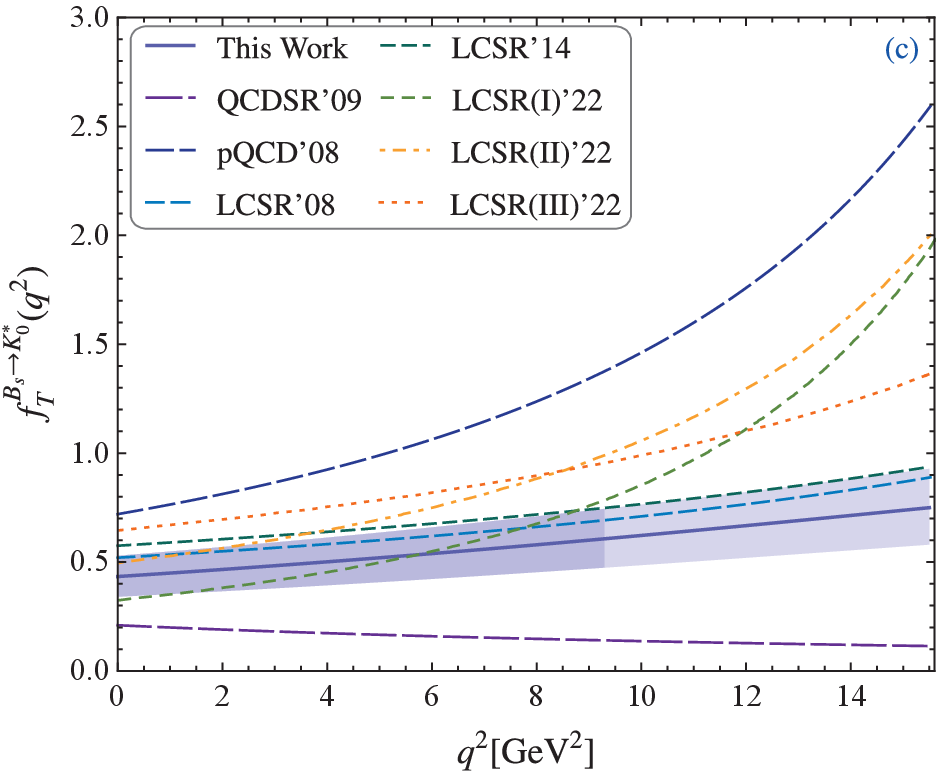}\\
~\includegraphics[width=0.32\textwidth]{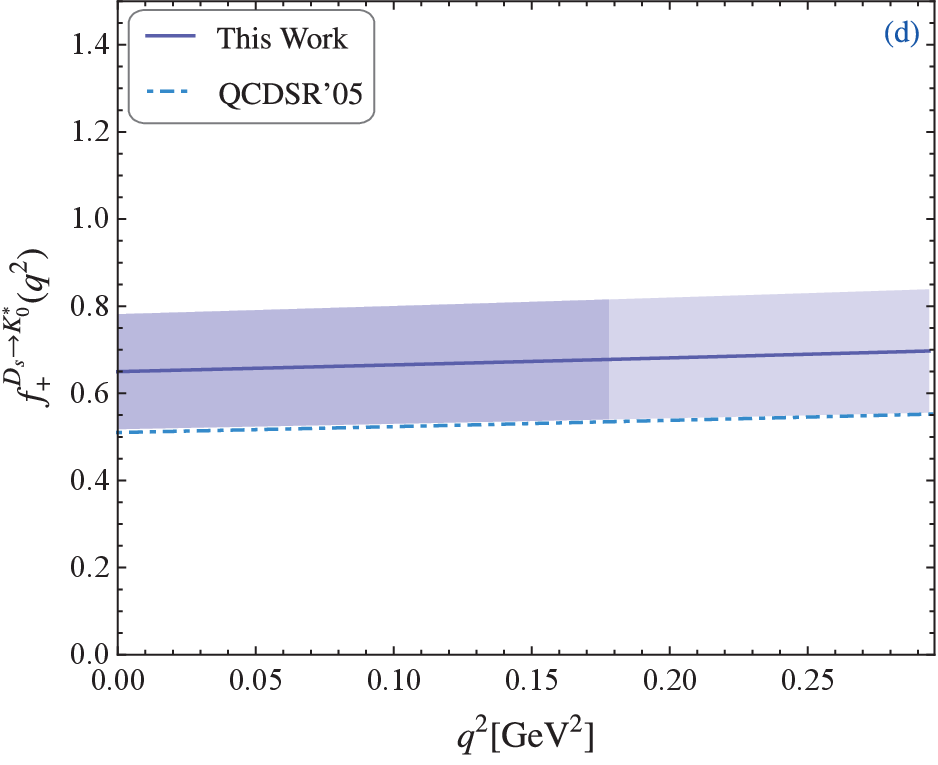}~\includegraphics[width=0.32\textwidth]{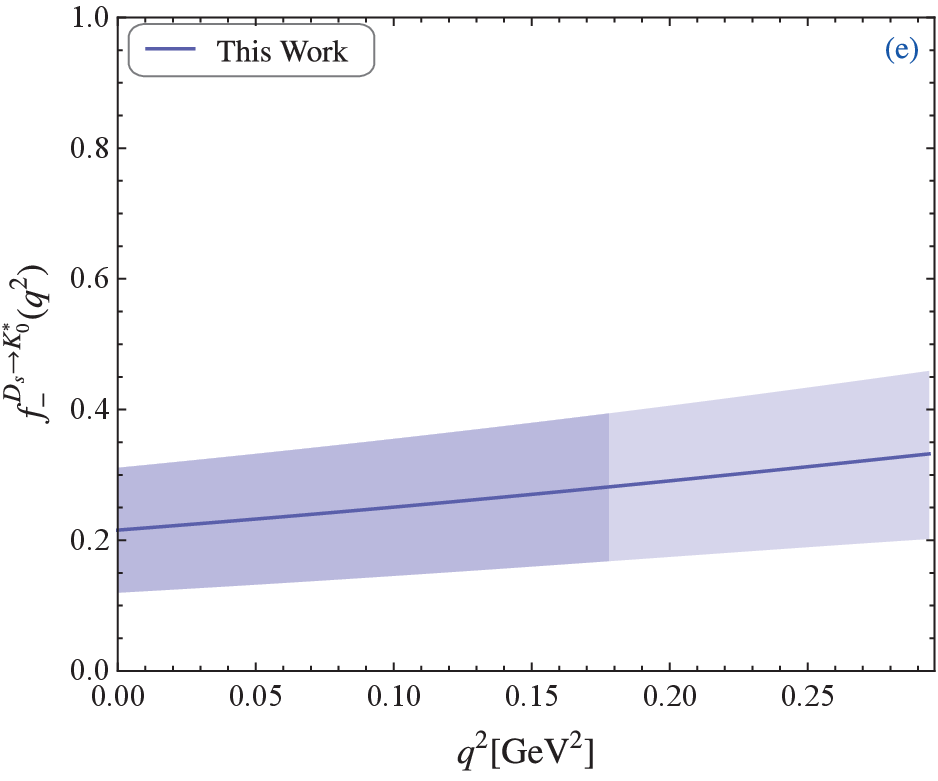}
~\includegraphics[width=0.32\textwidth]{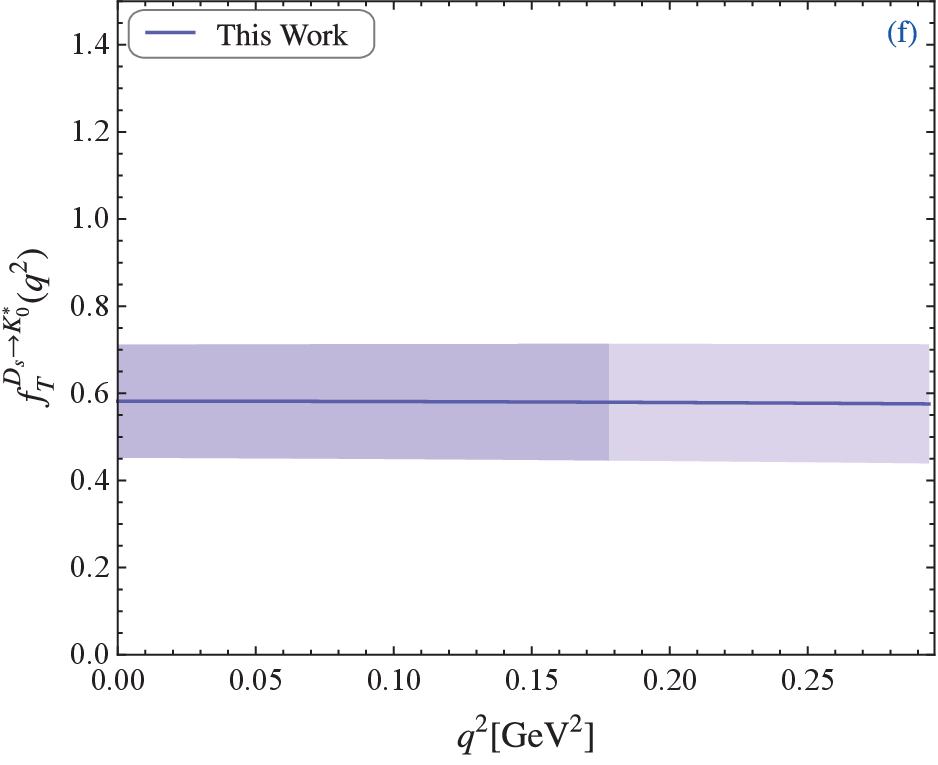}
\end{center}
\caption{Behaviors of the TFFs for the semileptonic $B_s, D_s \to K_0^\ast(1430)$ decays in whole $q^2$ region. The solid lines and shaded bands indicate the central values and uncertainties of our predictions, respectively. The dark parts and light parts in shaded bands are for the direct LCSR calculations and extrapolation results, respectively. We also shown other theoretical predictions such as RQM~\cite{Faustov:2013ima}, pQCD~\cite{Li:2008tk, Chen:2021oul}, QCDSR~\cite{Yang:2005bv, Ghahramany:2009zz}, LCSR~\cite{Khosravi:2022fzo, Wang:2008da, Wang:2014vra} for comparison.}
\label{fig:TFF}
\end{figure*}
In order to calculate the branching fractions of the semileptonic decays $B_s,D_s\to K_0^\ast(1430) \ell \nu_\ell$, one should extrapolate the results with LCSRs to the whole $q^2$ region, i.e., $q^2 \in [0, (m_{H_{q_1}} - m_{K_0^\ast})^2$]. In this work, we adopt the usual pole model parametrization~\cite{Li:2008tk, Chen:2021oul, Ghahramany:2009zz, Khosravi:2022fzo}
\begin{align}
f_i(q^2) = \frac{f_i(0)}{1 - a_i (q^2/m_{H_{q_1}}^2) + b_i (q^2/m_{H_{q_1}}^2)^2},
\end{align}
with $i = +,-,T$. The values of the extrapolation parameters $a_i$ and $b_i$ corresponding to the $B_s, D_s \to K_0^\ast(1430)$ TFFs are exhibited in Table~\ref{table:TFFsExt}. The behaviors of the $B_s, D_s \to K_0^\ast(1430)$ TFFs in whole $q^2$ region can be determined, and which are shown in Fig.~\ref{fig:TFF}. In Fig.~\ref{fig:TFF}, the solid lines are our central values and the shaded bands stand for the uncertainties. In particular, the dark parts are for the LCSR predictions, and the light parts are for the extrapolation results, respectively. Otherwise, the corresponding results calculated with RQM~\cite{Faustov:2013ima}, pQCD~\cite{Li:2008tk, Chen:2021oul}, QCDSR~\cite{Yang:2005bv, Ghahramany:2009zz}, LCSR~\cite{Khosravi:2022fzo, Wang:2008da, Wang:2014vra} are also shown in Fig.~\ref{fig:TFF} for comparison. From Fig.~\ref{fig:TFF} one can find that, our prediction for $f_+^{B_s\to K_0^\ast}(q^2)$ is consistent with the LCSR predictions in Refs.~\cite{Wang:2008da, Wang:2014vra} in entire $q^2$ region, and consists with pQCD calculation in Ref.~\cite{Chen:2021oul} in small recoil region. For $f_-^{B_s\to K_0^\ast}(q^2)$, our result is consistent with the QCDSR estimation of Ref.~\cite{Ghahramany:2009zz} in $q^2 \in [0, 9~{\rm GeV}^2]$. Meanwhile, our prediction for $f_T^{B_s\to K_0^\ast}(q^2)$ is consistent with the LCSR computation in Ref.~\cite{Wang:2008da} in whole $q^2$ region within the error range. In addition, some other predictions for $B_s\to K_0^\ast(1430)$ TFFs in the literature differ significantly from our results in large $q^2$ region, which requires future lattice QCD calculations near the small recoil point for judgement.

\begin{figure*}[htb]
\begin{center}
\includegraphics[width=0.43\textwidth]{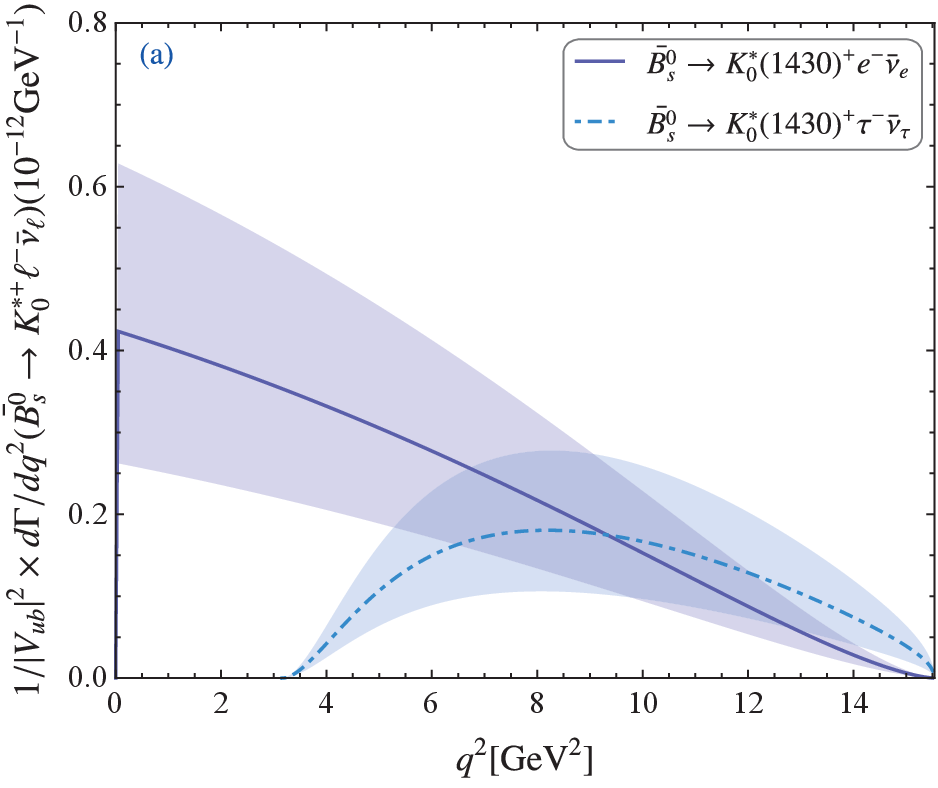}~~~~\includegraphics[width=0.43\textwidth]{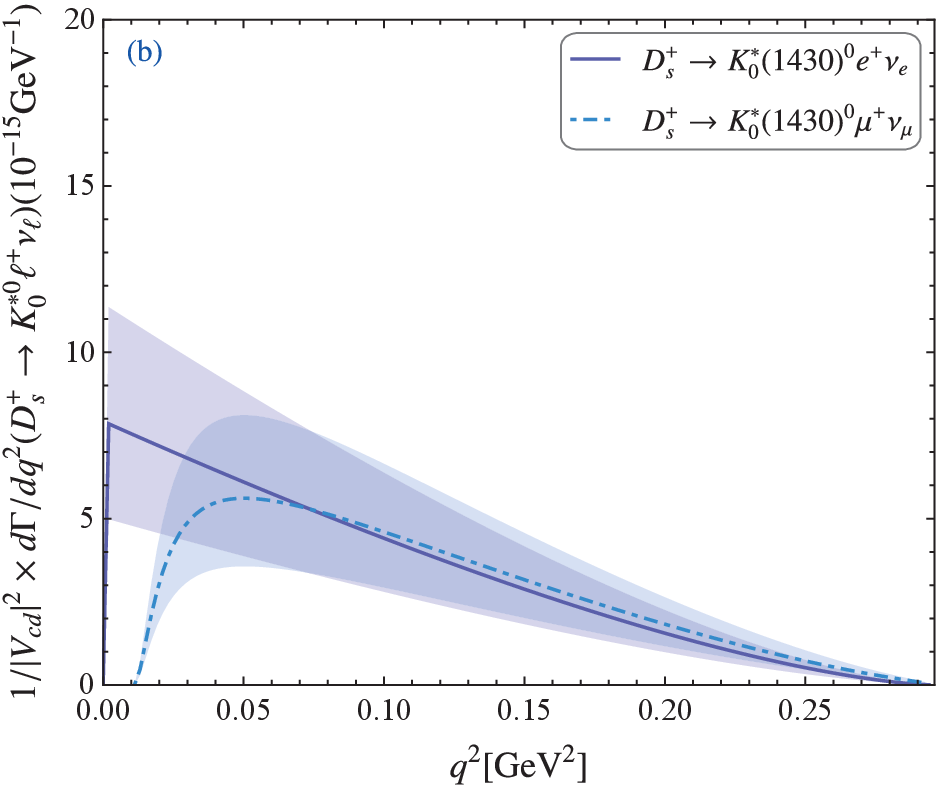}
\end{center}
\caption{Differential decay ratios of the semileptonic decays $\overline{B}_s^0 \to K_0^\ast(1430)^+ \ell^- \bar{\nu}_\ell$ with $\ell = e, \tau$ and $D_s^+ \to K_0^\ast(1430)^0 \ell^+ \nu_\ell$ with $\ell = e, \mu$, respectively. The solid and dashed lines are for the central values and the shaded bands are for the corresponding errors.}
\label{fig:DDW}
\end{figure*}

\begin{table*}[t]
\footnotesize
\begin{center}
\caption{Branching fractions $(\times 10^4)$ of the semileptonic decays $\overline{B}_{s}^{0} \to K_0^\ast(1430)^{+}\ell^{-} \bar{\nu}_{\ell}$ with $\ell = e, \mu$ and $\tau$, respectvely.}
\label{table:BRB}
\begin{tabular}{l c c c}
\hline\hline
 & ~~~~~~~~~~$\mathcal{B}(\overline{B}_s^0\to K_0^\ast(1430)^+ e^- \bar{\nu}_e)$~~~~~~~~~~ & ~~~~~~~~~~$\mathcal{B}(\overline{B}_s^0\to K_0^\ast(1430)^+ \mu^- \bar{\nu}_\mu)$~~~~~~~~~~ & ~~~~~~~~~~$\mathcal{B}(\overline{B}_s^0\to K_0^\ast(1430)^+ \tau^- \bar{\nu}_\tau)$~\\
This work~ & ~$1.13_{-0.51}^{+0.74}$~ & ~$1.13_{-0.51}^{+0.74}$~ & $0.50_{-0.25}^{+0.40}$~\\
Ref.~\cite{Faustov:2013ima}~ & ~$0.71^{+0.14}_{-0.14}$~ & ~$0.71^{+0.14}_{-0.14}$~ & ~$0.21^{+0.04}_{-0.04}$~ \\
Ref.~\cite{Li:2008tk}~ & ~$2.45^{+1.77}_{-1.05}$~ & ~$2.45^{+1.77}_{-1.05}$~ & ~$1.09^{+0.82}_{-0.47}$~ \\
Ref.~\cite{Khosravi:2022fzo}(I)~ & ~$-$~ & ~$0.99^{+0.89}_{-0.37}$~ & $0.49^{+0.33}_{-0.17}$~\\
Ref.~\cite{Khosravi:2022fzo}(II)~ & ~$-$~ & ~$1.67^{+1.32}_{-0.53}$~ & $0.71^{+0.57}_{-0.26}$~\\
Ref.~\cite{Khosravi:2022fzo}(III)~ & ~$-$~ & ~$1.90^{+1.48}_{-0.63}$~ & $0.65^{+0.55}_{-0.24}$~\\
Ref.~\cite{Wang:2008da}~ & ~$1.3^{+1.3}_{-0.4}$~ & ~$1.3^{+1.2}_{-0.4}$~ & ~$0.52^{+0.57}_{-0.18}$~  \\
Ref.~\cite{Wang:2014upa}~ & ~$1.27^{+0.36}_{-0.36}$~ & ~$1.27^{+0.36}_{-0.36}$~ & ~$0.54^{+0.16}_{-0.16}$~ \\
\hline\hline
\end{tabular}
\end{center}
\end{table*}
\begin{table}[t]
\footnotesize
\begin{center}
\caption{Branching fractions $(\times 10^4)$ of the semileptonic decays $D_{s}^{+} \to K_0^\ast(1430)^{0} \ell^{+} \nu_{\ell}$ with $\ell = e$ and $\mu$, respectively.}
\label{table:BRD}
\begin{tabular}{l l l}
\hline\hline
 &$\mathcal{B}(D_s^+\to K_0^\ast(1430)^0 e^+ \nu_e)$ &$\mathcal{B}(D_s^+\to K_0^\ast(1430)^0 \mu^+ \nu_\mu)$  \\
This work &~~~~~~~~~~~$0.36^{+0.19}_{-0.14}$ &~~~~~~~~~~~$0.31^{+0.16}_{-0.12}$  \\
Ref.~\cite{Yang:2005bv} &~~~~~~~~~~~$0.24^{+0.22}_{-0.15}$ &~~~~~~~~~~~$0.24^{+0.22}_{-0.15}$ \\
\hline\hline
\end{tabular}
\end{center}
\end{table}

Then, we can calculate the differential decay widths of the semileptonic decays $B_s,D_s\to K_0^\ast(1430) \ell \nu_\ell$ with the following formula~\cite{Yang:2005bv, Sun:2010nv}
\begin{align}
&\frac{d\Gamma}{dq^2} (H_{q_1}\to K_0^\ast \ell \nu_\ell) = \frac{G_F^2 |V_{Qq_2}|^2}{192\pi^3 m_{H_{q_1}}^3} \frac{q^2 - m_\ell^2}{(q^2)^2} \sqrt{\frac{(q^2-m_\ell^2)^2}{q^2}} \nonumber\\
&\times \sqrt{\frac{(m_{H_{q_1}}^2 - m_{K_0^\ast}^2 - q^2)^2}{4q^2} - m_{K_0^\ast}^2} \Big\{ (m_\ell^2 + 2q^2) \Big[ q^2 - (m_{H_{q_1}} \nonumber\\
&- m_{K_0^\ast})^2 \Big] \Big[ q^2 - (m_{H_{q_1}} + m_{K_0^\ast})^2 \Big] f_+^2(q^2) + 3m_\ell^2 (m_{H_{q_1}}^2 \nonumber\\
&- m_{K_0^\ast}^2)^2 \Big[ f_+(q^2) + \frac{q^2}{m_{H_{q_1}} - m_{K_0^\ast}} f_-(q^2) \Big]^2 \Big\}.
\label{eq:DDW}
\end{align}
In calculation, we take the fermi coupling constant $G_{F} = 1.166 \times 10^{-5}~\rm{GeV^{-2}}$, CKM Matrix elements $|V_{ub}| = (3.82 \pm 0.20) \times 10^{-3}$, $|V_{cd}| = 0.221 \pm 0.004$, $m_\ell$ denotes the lepton mass and $m_e = 0.511 \times 10^{-3}~{\rm MeV}$, $m_\mu = 105.658 \times 10^{-3}~{\rm MeV}$ and $m_\tau = 1776.86\times10^{-3}~{\rm MeV}$~\cite{Workman:2022ynf}. The resulted differential decay ratios for semileptonic decays $\overline{B}_s^0 \to K_0^\ast(1430)^+ \ell^- \bar{\nu}_\ell$ with $\ell = e, \tau$ and $D_s^+ \to K_0^\ast(1430)^0 \ell^+ \nu_\ell$ with $\ell = e, \mu$ versus $q^2$ are shown in Fig.~\ref{fig:DDW}, where the solid and dashed lines are for the central values and the shaded bands are for the corresponding errors.

Integrating Eq.~\eqref{eq:DDW} over $q^2$ in the region $m_\ell^2 \le q^2 \le (m_{H_{q_1}} - m_{K_0^\ast})^2$, and using the heavy meson mean lifetimes $\tau_{\overline{B}_s^0} = 1.520(5) \times 10^{-12}~s$ and $\tau_{D_s^+} = 0.504(4) \times 10^{-12}~s$~\cite{Workman:2022ynf}, the branching fractions of the semileptonic decays $\overline{B}_{s}^{0} \to K_0^\ast(1430)^{+}\ell^{-} \bar{\nu}_{\ell}$ with $\ell = e, \mu, \tau$ and $D_{s}^{+} \to K_0^\ast(1430)^{0} \ell^{+} \nu_{\ell}$ with $\ell = e, \mu$ can be obtained and are exhibited in Table~\ref{table:BRB} and Table~\ref{table:BRD}, respectively. As a comparison, the corresponding branching fractions obtained in Refs.~\cite{Faustov:2013ima, Li:2008tk, Yang:2005bv, Khosravi:2022fzo, Wang:2008da, Wang:2014upa} are also exhibited in Table~\ref{table:BRB} and Table~\ref{table:BRD}.

\section{SUMMARY}

The semileptonic $B_s, D_s \to K_0^\ast(1430)$ decays can provide another option for testing standard model beyond the semileptonic progresses with pseudoscalar mesons in the final states. In which, the $B_s, D_s \to K_0^\ast(1430)$ TFFs are the key objects, and whose accuracy mainly depends on the main error source, $\phi_{2;K_0^\ast}(x,\mu)$, the $K_0^\ast(1430)$ leading-twist DA. Motivated by this, we have studied the $K_0^\ast(1430)$ leading-twist DA and the semileptonic $B_s, D_s \to K_0^\ast(1430)$ decays in detail in this article. Our work is based on the scenario that the $K_0^\ast(1430)$ is viewed as the ground state of $s\bar{q}$ and $q\bar{s}$.

The $K_0^\ast(1430)$ leading-twist DA has studied following the scheme proposed in Ref.~\cite{Zhong:2021epq} at the first time. The $\xi$-moments are calculated with the QCDSRs in the framework of BFT by taking the $SU_f(3)$ symmetry breaking into account. Considering the fact that the zeroth $\xi$-moment of the $K_0^\ast(1430)$ twist-3 DA, $\langle\xi^0_p\rangle_{3;K_0^\ast}$, cannot be normalized in whole Borel region, a more reasonable and accurate sum rule formula for $n$th $\xi$-moments of $K_0^\ast(1430)$ leading-twist DA, i.e., Eq.~\eqref{SRxin}, has been suggested. The values of the first ten $\xi$-moments, $\langle\xi^n\rangle_{2;K_0^\ast} (n = 1,2,\cdots,10)$, have been calculated and exhibited in Table~\ref{table:xin1}. On the other hand, a new LCHO model has been established at the first time to describe the behavior of the $K_0^\ast(1430)$ leading-twist DA. By fitting the resulted $\xi$-moments shown in Table~\ref{table:xin1} with this LCHO model via the least squares method, the behavior of the $K_0^\ast(1430)$ leading-twist DA has been determined. The fitted model parameters at scale $\mu = 1~{\rm GeV}$, $1.4~{\rm GeV}$ and $3~{\rm GeV}$ have been displayed in Table~\ref{table:DA Parameters}, the predicted curve of $\phi_{2;K_0^\ast}(x,\mu)$ at scale $\mu = 1~{\rm GeV}$ is shown in Fig.~\ref{fig:DA}, respectively.

Then, we have calculated the $B_s, D_s\to K_0^\ast(1430)$ TFFs $f_{\pm, T}(q^2)$ with LCSR method. The values of those TFFs at the large recoil point have been given in Table~\ref{table:TFFsB} and Table~\ref{table:TFFsD}. After extrapolating the LCSR results of $B_s, D_s\to K_0^\ast(1430)$ TFFs to the whole $q^2$ region (the corresponding behaviors have been shown in Fig.~\ref{fig:TFF}), the differential decay ratios and branching fractions of the semileptonic decays $B_s, D_s\to K_0^\ast(1430) \ell \nu_\ell$ have been obtained and shown in Fig.~\ref{fig:DDW}, Table~\ref{table:BRB} and Table~\ref{table:BRD} respectively.

In addiction, we also perform the numerical calculations for the $K_0^\ast(1430)$ leading-twist DA and $B_s, D_s \to K_0^\ast(1430)$ TFFs and branching fractions in the framework of S1, i.e., the $K_0^\ast(1430)$ is assumed to be the excited state, the corresponding results are shown in Appendix B.\\

{\bf Acknowledgments}:
This work was supported in part by the National Natural Science Foundation of China under Grant No.12265009, No.12265010, No.12175025 and No.12147102, the Project of Guizhou Provincial Department of Science and Technology under Grant No.ZK[2021]024, No.ZK[2023]142, the Project of Guizhou Provincial Department of Education under Grant No.KY[2021]030, and by the Chongqing Graduate Research and Innovation Foundation under Grant No. ydstd1912. \\

\appendix

\section{Specific expressions of some condensate terms in sum rules of $\langle\xi^n\rangle_{2;K_0^\ast} \times \langle\xi^n_p\rangle_{3;K_0^\ast}$}

\begin{widetext}
\begin{align}
\hat{I}_{\langle G^2\rangle}(M^2) &= \frac{\langle\alpha_s G^2\rangle}{(M^2)^2} ((-1)^n m_s - m_u) \frac{1}{48\pi} \Big\{ -12(-1)^n n\Big( -\ln \frac{M^2}{\mu^2} \Big) + 6(-1)^n(n+2) + \theta(n-1) \Big[ -4(-1)^n n \Big( -\ln \frac{M^2}{\mu^2} \Big) \nonumber\\
&- 3\widetilde{\psi}_3(n) \Big] + \theta(n-2) \Big[ -(8n+3) \widetilde{\psi}_2(n) + (-1)^n(4n+9) + 7 + \frac{6}{n} \Big] \Big\}, \\
\hat{I}_{\langle G^3\rangle}(M^2) &= \frac{\langle g_s^3fG^3\rangle}{(M^2)^3} ((-1)^n m_s - m_u) \frac{1}{384\pi^2} \Big\{ \delta^{n1} \Big[ -24 \Big( -\ln \frac{M^2}{\mu^2} \Big) + 84 \Big] + \theta(n-1) \Big[ 4(-1)^n n(3n-5) \Big( -\ln \frac{M^2}{\mu^2} \Big) \nonumber\\
&- 2(-1)^n (2n^2 + 5n - 13) \Big] + \theta(n-2) \Big[ 24(-1)^n n^2 \Big( -\ln \frac{M^2}{\mu^2} \Big) + 2n(n-4) \widetilde{\psi}_2(n) - 17(-1)^n n^2 - 55(-1)^n n \nonumber\\
&- 6(-1)^n + 6 \Big] + \theta(n-3) \Big[ -2n(8n-1) \widetilde{\psi}_1(n) + \frac{1}{n-1} \Big( -19(-1)^n n^3 + 16(1+(-1)^n)n^2 + 3(2+(-1)^n)n \nonumber\\
&+ 6 \Big) \Big] \Big\}, \\
\hat{I}_{\langle q^4\rangle}(M^2) &= \frac{(2+\kappa^2)\langle g_s^2\bar{u}u\rangle^2}{(M^2)^3} ((-1)^n m_s - m_u) \frac{1}{3888\pi^2} \Big\{ 8(-1)^n n \Big( -\ln \frac{M^2}{\mu^2} \Big) - 4(-1)^n (n+5) + \delta^{n0} \Big[ -24\Big( -\ln \frac{M^2}{\mu^2} \Big) \nonumber\\
&- 148 \Big] + \delta^{n1} \Big[ 128 \Big( -\ln \frac{M^2}{\mu^2} \Big) - 692 \Big] + \theta(n-1) \Big[ -8 \Big( 6(-1)^n n^2 + (-25+9(-1)^n)n - 6 (1+(-1)^n) \Big) \Big( -\ln \frac{M^2}{\mu^2} \Big) \nonumber\\
&- 4n \widetilde{\psi}_3(n) + 2 \Big( 6(-1)^n n^2 + (-47+49(-1)^n)n + 57(-1)^n - 151 - \frac{24}{n} (1+(-1)^n) \Big) \Big] + \theta(n-2) \Big[ -4 \Big( 66(-1)^n n^2 \nonumber\\
&- 34(-1)^n n + 15(1+(-1)^n) \Big) \Big( -\ln \frac{M^2}{\mu^2} \Big) + 2\Big( -6n^2 + (-21+50(-1)^n)n + 12(1+(-1)^n) \Big) \widetilde{\psi}_2(n) \nonumber\\
&+ \frac{1}{n(n-1)} \Big( 231(-1)^n n^4 + (-94+426(-1)^n)n^3 + (116-985(-1)^n)n^2 + (98+328(-1)^n)n - 60(1 \nonumber\\
&+(-1)^n) \Big) \Big] + \theta(n-3) \Big[ \Big( 144n^2 - 74n + 30(1+(-1)^n) \Big) \widetilde{\psi}_1(n) + \frac{1}{n-1} \Big( 169(-1)^n n^3 - 12(12+17(-1)^n)n^2 \nonumber\\
&+ (10+111(-1)^n)n - 2(63+38(-1)^n) \Big) \Big] \Big\}, \\
\hat{I}_{\langle G^2\rangle}^{m_s^3}(M^2) &= \frac{\langle\alpha_sG^2\rangle}{(M^2)^3} m_s^3 \frac{1}{24\pi} \Big\{ -2\delta^{n0} - 10\delta^{n1} \Big[ \Big( -\ln \frac{M^2}{\mu^2} \Big) - \frac{17}{5} \Big] + \theta(n-1) \Big[ 6n \Big( -\ln \frac{M^2}{\mu^2} \Big) - 3(n+3) \Big] + \theta(n-2) \nonumber\\
&\times \Big[ -2n(7n-2) \Big( -\ln \frac{M^2}{\mu^2} \Big) + 3(-1)^n n \widetilde{\psi}_2(n) + 9n^2 + 24n - 3(-1)^n - 2 \Big] + \theta(n-3) \Big[ (-1)^n n(7n-2) \widetilde{\psi}_1(n) \nonumber\\
&- \frac{1}{n-1} \Big( -7n^3 + (5+7(-1)^n)n^2 + (2+(-1)^n)n + 2(-1)^n \Big) \Big] \Big\}, \\
\hat{I}_{\langle G^3\rangle}^{m_s^3}(M^2) &= \frac{\langle g_s^3fG^3\rangle}{(M^2)^4} m_s^3 \frac{1}{6912\pi^2} \Big\{ -24\delta^{n0} - 288\delta^{n1} + \delta^{n2} \Big[ 2352 \Big( -\ln \frac{M^2}{\mu^2} \Big) - 10188 \Big] + \theta(n-2) \Big[ 144n(n-1) \nonumber\\
&\times (2n-5) \Big( -\ln \frac{M^2}{\mu^2} \Big) - 9(8n^3 + 21n^2 - 205n + 90) \Big] + \theta(n-3) \Big[ 24n(n-1)(36n-23) \Big( -\ln \frac{M^2}{\mu^2} \Big) \nonumber\\
&- 144(-1)^n n(n-1)(n-2) \widetilde{\psi}_1(n) - 36(-1)^n n(n-1) \widetilde{\psi}_5(n) - 72(-1)^n n(n-1) \widetilde{\psi}_6(n) + \frac{2}{n-2} \Big( -505n^4 \nonumber\\
& +(281+72(-1)^n)n^3 - 2(-1274+99(-1)^n)n^2 + (-2417 + 135(-1)^n)n + 90(-1)^n + 474) \Big) \Big] + \theta(n-4) \nonumber\\
&\times \Big[ 12(-1)^n n(n-1)(36n-23) \widetilde{\psi}_4(n) + \frac{1}{n-2} \Big( -566n^4 + (1987-432(-1)^n)n^3 + (-1999+492(-1)^n)n^2 \nonumber\\
&+ (578-354(-1)^n)n - 156(-1)^n \Big) \Big] \Big\}, \\
\hat{I}_{\langle q^4\rangle}^{m_s^3}(M^2) &= \frac{(2+\kappa^2)\langle g_s^2\bar{u}u\rangle^2}{(M^2)^4} m_s^3 \frac{1}{23328\pi^2} \Big\{ \delta^{n0} \Big[ 1368 \Big( -\ln \frac{M^2}{\mu^2} \Big) + 1512 \Big] + \delta^{n1} \Big[ 1032 \Big( -\ln \frac{M^2}{\mu^2} \Big) - 1272 \Big] + \delta^{n2} \nonumber\\
&\times \Big[ -3504 \Big( -\ln \frac{M^2}{\mu^2} \Big) + 13980 \Big] - 48(-1)^n n \theta(n-1) + \theta(n-2) \Big[ -12 \Big( -8n^3 + 86n^2 + 4(-39+25(-1)^n)n \nonumber\\
&- 9(1+(-1)^n) \Big) \Big( -\ln \frac{M^2}{\mu^2} \Big) + \frac{3}{n(n-1)} \Big( -48n^5 + 429n^4 + (70+204(-1)^n)n^3 + (-1473+596(-1)^n)n^2 \nonumber\\
&+ (622-472(-1)^n)n + 36(1+(-1)^n) \Big) \Big] + \theta(n-3) \Big[ 24 \Big( -76n^3 + 177n^2 - 147n + 24(1+(-1)^n) \Big) \nonumber\\
&\times \Big( -\ln \frac{M^2}{\mu^2} \Big) - 6 \Big( 8(-1)^n n^3 - 86(-1)^n n^2 + 4(-25+39(-1)^n)n + 9(1+(-1)^n) \Big) \widetilde{\psi}_1(n) - \frac{2}{n(n-1)(n-2)} \nonumber\\
&\times \Big( -943n^6 - 3(-519+8(-1)^n)n^5 + (6041-24(-1)^n)n^4 - 3(5720-117(-1)^n)n^3 + (15359+363(-1)^n)n^2 \nonumber\\
&- 6(857+259(-1)^n)n + 576(1+(-1)^n) \Big) \Big] + \theta(n-4) \Big[ -12 \Big( 76(-1)^n n^3 - 177(-1)^n n^2 + 147(-1)^n n \nonumber\\
&- 24(1+(-1)^n) \Big) \widetilde{\psi}_4(n) + \frac{1}{(n-1)(n-2)} \Big( 1106n^5 + (-5469+912(-1)^n)n^4 + (10358-2580(-1)^n)n^3 \nonumber\\
&+ (-9669+3294(-1)^n)n^2 + (4250-798(-1)^n)n - 12(48+73(-1)^n) \Big) \Big] \Big\},
\end{align}
where
\begin{align}
\widetilde{\psi}_1(n) &= \psi\Big(\frac{n}{2}\Big) - \psi\Big(\frac{n-1}{2}\Big) - (-1)^n \ln 4, \nonumber\\
\widetilde{\psi}_2(n) &= \psi\Big(\frac{n+1}{2}\Big) - \psi\Big(\frac{n}{2}\Big) + (-1)^n \ln 4, \nonumber\\
\widetilde{\psi}_3(n) &= \psi\Big(\frac{n}{2}+1\Big) - \psi\Big(\frac{n+1}{2}\Big) - (-1)^n \ln 4, \nonumber\\
\widetilde{\psi}_4(n) &= \psi\Big(\frac{n-1}{2}\Big) - \psi\Big(\frac{n}{2}-1\Big) + (-1)^n \ln 4, \nonumber\\
\widetilde{\psi}_5(n) &= \psi\Big(\frac{n+1}{2}\Big) - \psi\Big(\frac{n}{2}-1\Big) - \ln 4, \nonumber\\
\widetilde{\psi}_6(n) &= \psi\Big(n-2\Big) - \psi\Big(\frac{n}{2}-1\Big) + (-1)^n \ln 4
\end{align}
with the digamma function $\psi(n)$.
\end{widetext}

\section{Numerical results corresponding to S1}

In S1, the $K_0^\ast(1430)$ is assumed to be the excited state corresponding ground state $\kappa$ (also known as $K_0^\ast(800)$). In this case, the new sum rule formula~\eqref{SRxin} is no longer applicable for calculating the $\xi$-moments of $K_0^\ast(1430)$ leading-twist DA. Then we adopt the method in Ref.~\cite{Cheng:2005nb}. That is, one can use Eq.~\eqref{SRxinxi0} to calculate the $\langle\xi^n\rangle_{2;K_0^\ast(1430)}$, but it needs to be modified. We abandoned considering the impact of the sum rule of $\langle\xi^0_p\rangle_{3;K_0^\ast}$ not being able to be normalized in the entire Borel parameter region and instead take $\langle\xi^0_p\rangle_{3;K_0^\ast} = 1$. Furthermore, the left side of Eq.~\eqref{SRxinxi0} should be replaced as follows,
\begin{align}
&\frac{\langle\xi^n\rangle_{2;K_0^\ast} \langle\xi^0_p\rangle_{3;K_0^\ast} m_{K_0^\ast} \bar{f}^2_{K_0^\ast}}{M^2 e^{m_{K_0^\ast}^2/M^2}} \nonumber\\
&\quad\quad \longrightarrow \frac{\langle\xi^n\rangle_{2;\kappa} m_{\kappa} \bar{f}^2_{\kappa}}{M^2 e^{m_{\kappa}^2/M^2}} + \frac{\langle\xi^n\rangle_{2;K_0^\ast} m_{K_0^\ast} \bar{f}^2_{K_0^\ast}}{M^2 e^{m_{K_0^\ast}^2/M^2}}. \nonumber
\end{align}

\begin{table*}[t]
\centering
\caption{Our predictions for the first three $\xi$-moments and Gegenbauer moments of $\kappa$ and $K_0^\ast(1430)$ leading-twist DAs at the scale $\mu = 1~{\rm GeV}$ in S1.}\label{table:xin_value_S1}
\begin{tabular}{l c c c c c c}
\hline\hline
$S$~& ~~~~~~$\langle\xi^1\rangle_{2;S}$~~~~~~ & ~~~~~~$\langle\xi^2\rangle_{2;S}$~~~~~~ & ~~~~~~$\langle\xi^3\rangle_{2;S}$~~~~~~ & ~~~~~~$a_1^{2;S}$~~~~~~ & ~~~~~~$a_2^{2;S}$~~~~~~ & ~~~~~~$a_3^{2;S}$~~~~~~  \\
\hline
$\kappa$               & $-0.454_{-0.064}^{+0.065}$ & $~~0.033_{-0.010}^{+0.009}$ & $-0.304_{-0.046}^{+0.047}$ & $-0.757_{-0.106}^{+0.109}$ & $~~0.096_{-0.028}^{+0.025}$ & $-0.575_{-0.100}^{+0.102}$ \\
$K_0^\ast(1430)$       & $-0.017^{+0.042}_{-0.023}$ & $-0.017^{+0.008}_{-0.010}$ & $-0.044^{+0.043}_{-0.036}$ & $-0.028^{+0.070}_{-0.038}$ & $-0.050^{+0.023}_{-0.029}$ & $-0.192^{+0.132}_{-0.139}$ \\
\hline\hline
\end{tabular}
\end{table*}

\begin{table}[t]
\footnotesize
\begin{center}
\caption{TFFs of the semileptonic decays $B_s\to K_0^\ast(1430)$ and $D_s\to K_0^\ast(1430)$ at large recoil point in S1.}~\label{table:TFFs_S1}
\begin{tabular}{l l l l}
\hline\hline
~Method~~~~~~ & $f_+^{B_s\to K_0^\ast}(0)$~~~~~~ & $f_-^{B_s\to K_0^\ast}(0)$~~~~~~ & $f_T^{B_s\to K_0^\ast}(0)$~ \\
\hline
~This work~  & $0.22_{-0.05}^{+0.05}$  & $-0.07_{-0.02}^{+0.02}$  & $0.18_{-0.04}^{+0.04}$  \\
~pQCD~\cite{Li:2008tk}~  & $-0.32^{+0.06}_{-0.07}$  & $-$  & $-0.41^{+0.08}_{-0.09}$  \\
~pQCD~\cite{Zhang:2010af}~  & $-0.30^{+0.03}_{-0.03}$  & $-$  & $-$   \\
~pQCD~\cite{Chen:2021oul}~  & $0.23^{+0.02}_{-0.02}$  & $-$  & $-$    \\
~LCSR~\cite{Sun:2010nv}~  & $0.10$  & $-0.10$  & $-$   \\
\hline\hline
~~~~~~ & $f_+^{D_s\to K_0^\ast}(0)$~~~~~ & $f_-^{D_s\to K_0^\ast}(0)$~~~~~ & $f_T^{D_s\to K_0^\ast}(0)$~ \\
\hline
~This work~  & $0.58_{-0.12}^{+0.12}$  & $0.28_{-0.09}^{+0.09}$  & $0.40_{-0.08}^{+0.08}$   \\
\hline\hline
\end{tabular}
\end{center}
\end{table}

\begin{table*}[t]
\footnotesize
\begin{center}
\caption{Branching fractions $(\times 10^4)$ of the semileptonic decays $\overline{B}_{s}^{0} \to K_0^\ast(1430)^{+}\ell^{-} \bar{\nu}_{\ell}$ with $\ell = e, \mu, \tau$ and $D_{s}^{+} \to K_0^\ast(1430)^{0} \ell^{+} \nu_{\ell}$ with $\ell = e, \mu$ in S1, respectvely.}~\label{table:BR_S1}
\begin{tabular}{l c c c}
\hline\hline
 & ~~~~~~~~~~$\mathcal{B}(\overline{B}_s^0\to K_0^\ast(1430)^+ e^- \bar{\nu}_e)$~~~~~~~~~~ & ~~~~~~~~~~$\mathcal{B}(\overline{B}_s^0\to K_0^\ast(1430)^+ \mu^- \bar{\nu}_\mu)$~~~~~~~~~~ & ~~~~~~~~~~$\mathcal{B}(\overline{B}_s^0\to K_0^\ast(1430)^+ \tau^- \bar{\nu}_\tau)$~\\
This work~ & ~$0.43_{-0.19}^{+0.29}$~ & ~$0.43_{-0.19}^{+0.29}$~ & $0.27_{-0.13}^{+0.20}$~\\
Ref.~\cite{Li:2008tk}~ & ~$-$~ & ~$0.77^{+0.37}_{-0.27}$~ & ~$0.35^{+0.17}_{-0.12}$~ \\
\hline\hline
&$\mathcal{B}(D_s^+\to K_0^\ast(1430)^0 e^+ \nu_e)$ &$\mathcal{B}(D_s^+\to K_0^\ast(1430)^0 \mu^+ \nu_\mu)$  \\
This work~ & $0.30^{+0.16}_{-0.12}$ & $0.26^{+0.13}_{-0.11}$  \\
\hline\hline
\end{tabular}
\end{center}
\end{table*}
In numerical calculation, we take $m_\kappa =845\pm17 {\rm MeV}$~\cite{Workman:2022ynf}, $\bar{f}_\kappa =340\pm20 {\rm MeV}$ at $\mu = 1~{\rm GeV}$~\cite{Cheng:2005nb}, and the continuum threshold parameters $s_\kappa =2.4~ {\rm GeV}^2$ and $s_{K_0^\ast(1430)} =6~ {\rm GeV}^2$~\cite{Cheng:2005nb}, respectively. By taking the Borel windows as $M^2 \in [3,4]~{\rm GeV}^2$, the first three $\xi$-moments and Gegenbauer moments of $\kappa$ and $K_0^\ast(1430)$ leading-twist DAs in S1 can be obtained, and are exhibited in Table~\ref{table:xin_value_S1}. Our predictions for $\langle\xi^1\rangle_{2;\kappa}$ and $\langle\xi^3\rangle_{2;\kappa}$ are consistent with the values of Ref.~\cite{Cheng:2005nb} within the error region (In Ref.~\cite{Cheng:2005nb}, $\langle\xi^1\rangle_{2;\kappa} = -0.55\pm 0.07$ and $\langle\xi^3\rangle_{2;\kappa} = -0.21\pm 0.05$). However, our predictions for moments of $K_0^\ast(1430)$ leading-twist DA are much less than the corresponding results in Ref.~\cite{Cheng:2005nb}, and are also much less than our predictions in S2 (see Table~\ref{table:xin_value}).

Based on the truncation form of the Gegenbauer series expansion for $\phi_{2;K_0^\ast}(x,\mu)$~\cite{Cheng:2005nb}, i.e.,
\begin{align}
\phi_{2;K_0^\ast}(x,\mu) &= 6x(1-x) \Big[ a_1^{2;K_0^\ast} C_1^{3/2}(2x-1) \nonumber\\
&+ a_2^{2;K_0^\ast} C_2^{3/2}(2x-1) + a_3^{2;K_0^\ast} C_3^{3/2}(2x-1) \Big],
\end{align}
the $B_s, D_s \to K_0^\ast(1430)$ TFFs and the corresponding branching fractions can be calculated. The corresponding numerical results are shown in Table~\ref{table:TFFs_S1} and Table~\ref{table:BR_S1}. As a comparison, the other predictions corresponding to S1 by pQCD~\cite{Li:2008tk, Zhang:2010af, Chen:2021oul} and LCSR~\cite{Sun:2010nv} are also shown in Table~\ref{table:TFFs_S1} and Table~\ref{table:BR_S1}. One can find that, our $f_+^{B_s\to K_0^\ast}(0)$ is consistent with the one in Ref.~\cite{Chen:2021oul}.


\begin{thebibliography}{99}

\bibitem{Workman:2022ynf}
R.~L.~Workman \textit{et al.} (Particle Data Group),
\textrm{Review of Particle Physics},
\href{https://doi.org/10.1093/ptep/ptac097}
{Prog. Theor. Exp. Phys. \textbf{2022}, 083C01 (2022)}.

\bibitem{Faustov:2013ima}
R.~N.~Faustov and V.~O.~Galkin,
\textrm{Charmless weak $B_s$ decays in the relativistic quark model},
\href{https://doi.org/10.1103/PhysRevD.87.094028}
{Phys. Rev. D \textbf{87}, no.9, 094028 (2013)}.
\href{https://arxiv.org/abs/1304.3255}
{[arXiv:1304.3255]}

\bibitem{Li:2008tk}
R.~H.~Li, C.~D.~Lu, W.~Wang and X.~X.~Wang,
\textrm{$B\to S$ Transition Form Factors in the PQCD approach},
\href{https://doi.org/10.1103/PhysRevD.79.014013}
{Phys. Rev. D \textbf{79}, 014013 (2009)}.
\href{https://arxiv.org/abs/0811.2648}
{[arXiv:0811.2648]}

\bibitem{Zhang:2010af}
Z.~Q.~Zhang,
\textrm{Branching ratio and CP asymmetry of $B_s \to K^{0*}(1430) \rho (\omega, \phi)$ decays in the perturbative QCD approach},
\href{https://doi.org/10.1103/PhysRevD.82.114016}
{Phys. Rev. D \textbf{82}, 114016 (2010)}.
\href{https://arxiv.org/abs/1106.0103}
{[arXiv:1106.0103]}

\bibitem{Chen:2021oul}
L.~Chen, M.~Zhao, Y.~Zhang and Q.~Chang,
\textrm{Study of $B_{u,d,s} \to K^\ast_0$ (1430)$P$ and $K^\ast_0 (1430)V$ decays within QCD factorization},
\href{https://doi.org/10.1103/PhysRevD.105.016002}
{Phys. Rev. D \textbf{105}, no.1, 016002 (2022)}.
\href{https://arxiv.org/abs/2112.00915}
{[arXiv:2112.00915]}

\bibitem{Yang:2005bv}
M.~Z.~Yang,
\textrm{Semileptonic decay of $B$ and $D\to K_0^\ast(1430)\bar{\ell}\nu$ from QCD sum rule},
\href{https://doi.org/10.1103/PhysRevD.73.079901}
{Phys. Rev. D \textbf{73}, 034027 (2006)}
[erratum: Phys. Rev. D \textbf{73}, 079901 (2006)].
\href{https://arxiv.org/abs/hep-ph/0509103}
{[arXiv:hep-ph/0509103]}

\bibitem{Ghahramany:2009zz}
N.~Ghahramany and R.~Khosravi,
\textrm{Analysis of the rare semileptonic decays of $B_s$ to $f_0(980)$ and $K_0^\ast(1430)$ scalar mesons in QCD sum rules},
\href{https://doi.org/10.1103/PhysRevD.80.016009}
{Phys. Rev. D \textbf{80}, 016009 (2009)}.

\bibitem{Khosravi:2022fzo}
R.~Khosravi,
\textrm{Semileptonic $B_s\to K_0^\ast(1430)$ transitions with the light-cone sum rules},
\href{https://doi.org/10.1103/PhysRevD.105.116027}
{Phys. Rev. D \textbf{105}, no.11, 116027 (2022)}.
\href{https://arxiv.org/abs/2203.09997}
{[arXiv:2203.09997]}

\bibitem{Wang:2008da}
Y.~M.~Wang, M.~J.~Aslam and C.~D.~Lu,
\textrm{Scalar mesons in weak semileptonic decays of B(s)},
\href{https://doi.org/10.1103/PhysRevD.78.014006}
{Phys. Rev. D \textbf{78}, 014006 (2008)}.
\href{https://arxiv.org/abs/0804.2204}
{[arXiv:0804.2204]}

\bibitem{Wang:2014vra}
Z.~G.~Wang,
\textrm{$B-S$ transition form-factors with the light-cone QCD sum rules},
\href{https://doi.org/10.1140/epjc/s10052-015-3282-3}
{Eur. Phys. J. C \textbf{75}, no.2, 50 (2015)}.
\href{https://arxiv.org/abs/1409.6449}
{[arXiv:1409.6449]}

\bibitem{Wang:2014upa}
Z.~G.~Wang,
\textrm{Semi-leptonic $B\to S$ decays in the standard model and in the universal extra dimension model},
\href{https://doi.org/10.1016/j.nuclphysb.2015.07.015}
{Nucl. Phys. B \textbf{898}, 431-447 (2015)}.
\href{https://arxiv.org/abs/1411.7961}
{[arXiv:1411.7961]}

\bibitem{Sun:2010nv}
Y.~J.~Sun, Z.~H.~Li and T.~Huang,
\textrm{$B_{(s)}\to S$ transitions in the light cone sum rules with the chiral current},
\href{https://doi.org/10.1103/PhysRevD.83.025024}
{Phys. Rev. D \textbf{83}, 025024 (2011)}.
\href{https://arxiv.org/abs/1011.3901}
{[arXiv:1011.3901]}

\bibitem{Han:2013zg}
H.~Y.~Han, X.~G.~Wu, H.~B.~Fu, Q.~L.~Zhang and T.~Zhong,
\textrm{Twist-3 Distribution Amplitudes of Scalar Mesons within the QCD Sum Rules and Its Application to the $B \to S$ Transition Form Factors},
\href{https://doi.org/10.1140/epja/i2013-13078-7}
{Eur. Phys. J. A \textbf{49}, 78 (2013)}
\href{https://arxiv.org/abs/1301.3978}
{[arXiv:1301.3978]}

\bibitem{Cheng:2002ai}
H.~Y.~Cheng,
\textrm{Hadronic $D$ decays involving scalar mesons},
\href{https://doi.org/10.1103/PhysRevD.67.034024}
{Phys. Rev. D \textbf{67}, 034024 (2003)}.
\href{https://arxiv.org/abs/hep-ph/0212117}
{[arXiv:hep-ph/0212117]}

\bibitem{Duplancic:2008ix}
G.~Duplancic, A.~Khodjamirian, T.~Mannel, B.~Melic and N.~Offen,
\textrm{Light-cone sum rules for $B\to\pi$ form factors revisited},
\href{https://doi.org/10.1088/1126-6708/2008/04/014}
{JHEP \textbf{04}, 014 (2008)}.
\href{https://arxiv.org/abs/0801.1796}
{[arXiv:0801.1796]}

\bibitem{Duplancic:2008tk}
G.~Duplancic and B.~Melic,
\textrm{$B,B_s\to K$ form factors: An Update of light-cone sum rule results},
\href{https://doi.org/10.1103/PhysRevD.78.054015}
{Phys. Rev. D \textbf{78}, 054015 (2008)}.
\href{https://arxiv.org/abs/0805.4170}
{[arXiv:0805.4170]}

\bibitem{Cheng:2005nb}
H.~Y.~Cheng, C.~K.~Chua and K.~C.~Yang,
\textrm{Charmless hadronic $B$ decays involving scalar mesons: Implications to the nature of light scalar mesons},
\href{https://doi.org/10.1103/PhysRevD.73.014017}
{Phys. Rev. D \textbf{73}, 014017 (2006)}.
\href{https://arxiv.org/abs/hep-ph/0508104}
{[arXiv:hep-ph/0508104]}

\bibitem{Weinstein:1982gc}
J.~D.~Weinstein and N.~Isgur,
\textrm{Do Multi-Quark Hadrons Exist?}
\href{https://doi.org/10.1103/PhysRevLett.48.659}
{Phys. Rev. Lett. \textbf{48}, 659 (1982)}.

\bibitem{Jaffe:1976ig}
R.~L.~Jaffe,
\textrm{Multi-Quark Hadrons. 1. The Phenomenology of (2 Quark 2 anti-Quark) Mesons},
\href{https://doi.org/10.1103/PhysRevD.15.267}
{Phys. Rev. D \textbf{15}, 267 (1977)}.

\bibitem{Jaffe:1976ih}
R.~L.~Jaffe,
\textrm{Multi-Quark Hadrons. 2. Methods},
\href{https://doi.org/10.1103/PhysRevD.15.281}
{Phys. Rev. D \textbf{15}, 281 (1977)}.

\bibitem{Wang:2010pn}
Z.~G.~Wang,
\textrm{Analysis of the nonet scalar mesons as tetraquark states with new QCD sum rules},
\href{https://doi.org/10.1007/s10773-011-0929-1}
{Int. J. Theor. Phys. \textbf{51}, 507-517 (2012)}.
\href{https://arxiv.org/abs/1008.0974}
{[arXiv:1008.0974]}

\bibitem{Du:2004ki}
D.~S.~Du, J.~W.~Li and M.~Z.~Yang,
\textrm{Mass and decay constant of $I = 1/2$ scalar meson in QCD sum rule},
\href{https://doi.org/10.1016/j.physletb.2005.05.043}
{Phys. Lett. B \textbf{619}, 105-114 (2005)}.
\href{https://arxiv.org/abs/hep-ph/0409302}
{[arXiv:hep-ph/0409302]}

\bibitem{Lu:2006fr}
C.~D.~Lu, Y.~M.~Wang and H.~Zou,
\textrm{Twist-3 distribution amplitudes of scalar mesons from QCD sum rules},
\href{https://doi.org/10.1103/PhysRevD.75.056001}
{Phys. Rev. D \textbf{75}, 056001 (2007)}.
\href{https://arxiv.org/abs/hep-ph/0612210}
{[arXiv:hep-ph/0612210]}

\bibitem{Zhong:2022lmn}
T.~Zhong, Z.~H.~Zhu and H.~B.~Fu,
\textrm{Constraint of $\xi$-moments calculated with QCD sum rules on the pion distribution amplitude models},
\href{https://arxiv.org/abs/2209.02493}
{[arXiv:2209.02493]}.

\bibitem{Huang:1989gv}
T.~Huang and Z.~Huang,
\textrm{Quantum Chromodynamics in Background Fields},
\href{https://doi.org/10.1103/PhysRevD.39.1213}
{Phys. Rev. D \textbf{39}, 1213-1220 (1989)}.

\bibitem{Zhong:2021epq}
T.~Zhong, Z.~H.~Zhu, H.~B.~Fu, X.~G.~Wu and T.~Huang,
\textrm{Improved light-cone harmonic oscillator model for the pionic leading-twist distribution amplitude},
\href{https://doi.org/10.1103/PhysRevD.104.016021}
{Phys. Rev. D \textbf{104}, no.1, 016021 (2021)}.
\href{https://arxiv.org/abs/2102.03989}
{[arXiv:2102.03989]}

\bibitem{Zhong:2022ecl}
T.~Zhong, H.~B.~Fu and X.~G.~Wu,
\textrm{Investigating the ratio of CKM matrix elements $|V_{ub}|/|V_{cb}|$ from semileptonic decay $B_s^0\to K-\mu^+\nu_\mu$ and kaon twist-2 distribution amplitude},
\href{https://doi.org/10.1103/PhysRevD.105.116020}
{Phys. Rev. D \textbf{105}, no.11, 116020 (2022)}.
\href{https://arxiv.org/abs/2201.10820}
{[arXiv:2201.10820]}

\bibitem{Hu:2021lkl}
D.~D.~Hu, H.~B.~Fu, T.~Zhong, Z.~H.~Wu and X.~G.~Wu,
\textrm{$a_1(1260)$-meson longitudinal twist-2 distribution amplitude and the $D\rightarrow a_1(1260)\ell ^+\nu _\ell $ decay processes},
\href{https://doi.org/10.1140/epjc/s10052-022-10555-y}
{Eur. Phys. J. C \textbf{82}, no.7, 603 (2022)}.
\href{https://arxiv.org/abs/2107.02758}
{[arXiv:2107.02758]}

\bibitem{BHL}
S. J. Brodsky, T. Huang, and G. P. Lepage, in Particles and Fields-2, Proceedings of the Banff Summer Institute, Ban8; Alberta, 1981, edited by A. Z. Capri and A. N. Kamal (Plenum, New York, 1983), p. 143; G. P. Lepage, S. J. Brodsky, T. Huang, and P. B.Mackenize, ibid. , p. 83; T. Huang, in Proceedings of XXth International Conference on High Energy Physics, Madison, Wisconsin, 1980, edited by L. Durand and L. G Pondrom, AIP Conf. Proc. No. 69 (AIP, New York, 1981), p. 1000.

\bibitem{Zhong:2011rg}
T.~Zhong, X.~G.~Wu, H.~Y.~Han, Q.~L.~Liao, H.~B.~Fu and Z.~Y.~Fang,
\textrm{Revisiting the Twist-3 Distribution Amplitudes of $K$ Meson within the QCD Background Field Approach},
\href{https://doi.org/10.1088/0253-6102/58/2/16}
{Commun. Theor. Phys. \textbf{58}, 261-270 (2012)}.
\href{https://arxiv.org/abs/1109.3127}
{[arXiv:1109.3127]}

\bibitem{Zhong:2014jla}
T.~Zhong, X.~G.~Wu, Z.~G.~Wang, T.~Huang, H.~B.~Fu and H.~Y.~Han,
\textrm{Revisiting the Pion Leading-Twist Distribution Amplitude within the QCD Background Field Theory},
\href{https://doi.org/10.1103/PhysRevD.90.016004}
{Phys. Rev. D \textbf{90}, no.1, 016004 (2014)}
\href{https://arxiv.org/abs/1405.0774}
{[arXiv:1405.0774]}

\bibitem{Hu:2021zmy}
D.~D.~Hu, H.~B.~Fu, T.~Zhong, L.~Zeng, W.~Cheng and X.~G.~Wu,
\textrm{$\eta ^{(\prime )}$-meson twist-2 distribution amplitude within QCD sum rule approach and its application to the semi-leptonic decay $ D_s^+ \rightarrow \eta ^{(\prime )}\ell ^+ \nu _\ell $},
\href{https://doi.org/10.1140/epjc/s10052-021-09958-0}
{Eur. Phys. J. C \textbf{82}, no.1, 12 (2022)}
\href{https://arxiv.org/abs/2102.05293}
{[arXiv:2102.05293]}

\bibitem{Huang:1994dy}
T.~Huang, B.~Q.~Ma and Q.~X.~Shen,
\textrm{Analysis of the pion wave function in light cone formalism},
\href{https://doi.org/10.1103/PhysRevD.49.1490}
{Phys. Rev. D \textbf{49}, 1490-1499 (1994)}.
\href{https://arxiv.org/abs/hep-ph/9402285}
{[arXiv:hep-ph/9402285]}

\bibitem{Cao:1997hw}
F.~g.~Cao and T.~Huang,
\textrm{Large corrections to asymptotic $F_{\eta_c\gamma}$ and $F_{\eta_b\gamma}$ in the light cone perturbative QCD},
\href{https://doi.org/10.1103/PhysRevD.59.093004}
{Phys. Rev. D \textbf{59}, 093004 (1999)}.
\href{https://arxiv.org/abs/hep-ph/9711284}
{[arXiv:hep-ph/9711284 [hep-ph]]}

\bibitem{Huang:2004fn}
T.~Huang, X.~G.~Wu and X.~H.~Wu,
\textrm{Pion form-factor in the $k_T$ factorization formalism},
\href{https://doi.org/10.1103/PhysRevD.70.053007}
{Phys. Rev. D \textbf{70}, 053007 (2004)}.
\href{https://arxiv.org/abs/hep-ph/0404163}
{[arXiv:hep-ph/0404163]}

\bibitem{Wu:2005kq}
X.~G.~Wu and T.~Huang,
\textrm{Pion electromagnetic form-factor in the $k_T$ factorization formulae},
\href{https://doi.org/10.1142/S0217751X06032277}
{Int. J. Mod. Phys. A \textbf{21}, 901-904 (2006)}.
\href{https://arxiv.org/abs/hep-ph/0507136}
{[arXiv:hep-ph/0507136]}

\bibitem{Huang:2006wt}
T.~Huang and X.~G.~Wu,
\textrm{A Comprehensive Analysis on the Pion-Photon Transition Form Factor Involving the Transverse Momentum Corrections},
\href{https://doi.org/10.1142/S0217751X07036671}
{Int. J. Mod. Phys. A \textbf{22}, 3065-3086 (2007)}.
\href{https://arxiv.org/abs/hep-ph/0606135}
{[arXiv:hep-ph/0606135]}

\bibitem{Wu:2011gf}
X.~G.~Wu and T.~Huang,
\textrm{Constraints on the Light Pseudoscalar Meson Distribution Amplitudes from Their Meson-Photon Transition Form Factors},
\href{https://doi.org/10.1103/PhysRevD.84.074011}
{Phys. Rev. D \textbf{84}, 074011 (2011)}.
\href{https://arxiv.org/abs/1106.4365}
{[arXiv:1106.4365]}

\bibitem{Wu:2012kw}
X.~G.~Wu, T.~Huang and T.~Zhong,
\textrm{Information on the Pion Distribution Amplitude from the Pion-Photon Transition Form Factor with the Belle and BaBar Data},
\href{https://doi.org/10.1088/1674-1137/37/6/063105}
{Chin. Phys. C \textbf{37}, 063105 (2013)}.
\href{https://arxiv.org/abs/1206.0466}
{[arXiv:1206.0466]}

\bibitem{Huang:2013gra}
T.~Huang, X.~G.~Wu and T.~Zhong,
\textrm{Finding a way to determine the pion distribution amplitude from the experimental data},
\href{https://doi.org/10.1088/0256-307X/30/4/041201}
{Chin. Phys. Lett. \textbf{30}, 041201 (2013)}.
\href{https://arxiv.org/abs/1303.2301}
{[arXiv:1303.2301]}

\bibitem{Huang:2013yya}
T.~Huang, T.~Zhong and X.~G.~Wu,
\textrm{Determination of the pion distribution amplitude},
\href{https://doi.org/10.1103/PhysRevD.88.034013}
{Phys. Rev. D \textbf{88}, 034013 (2013)}.
\href{https://arxiv.org/abs/1305.7391}
{[arXiv:1305.7391]}

\bibitem{Zhong:2014fma}
T.~Zhong, X.~G.~Wu and T.~Huang,
\textrm{Heavy Pseudoscalar Leading-Twist Distribution Amplitudes within QCD Theory in Background Fields},
\href{https://doi.org/10.1140/epjc/s10052-015-3271-6}
{Eur. Phys. J. C \textbf{75}, no.2, 45 (2015)}.
\href{https://arxiv.org/abs/1408.2297}
{[arXiv:1408.2297]}

\bibitem{Zhong:2015nxa}
T.~Zhong, X.~G.~Wu and T.~Huang,
\textrm{The longitudinal and transverse distributions of the pion wave function from the present experimental data on the pion-photon transition form factor},
\href{https://doi.org/10.1140/epjc/s10052-016-4236-0}
{Eur. Phys. J. C \textbf{76}, no.7, 390 (2016)}.
\href{https://arxiv.org/abs/1510.06924}
{[arXiv:1510.06924]}

\bibitem{Zhong:2016kuv}
T.~Zhong, X.~G.~Wu, T.~Huang and H.~B.~Fu,
\textrm{Heavy Pseudoscalar Twist-3 Distribution Amplitudes within QCD Theory in Background Fields},
\href{https://doi.org/10.1140/epjc/s10052-016-4350-z}
{Eur. Phys. J. C \textbf{76}, no.9, 509 (2016)}.
\href{https://arxiv.org/abs/1604.04709}
{[arXiv:1604.04709]}

\bibitem{Zhang:2017rwz}
Y.~Zhang, T.~Zhong, X.~G.~Wu, K.~Li, H.~B.~Fu and T.~Huang,
\textrm{Uncertainties of the $B\rightarrow D$ transition form factor from the D-meson leading-twist distribution amplitude},
\href{https://doi.org/10.1140/epjc/s10052-018-5551-4}
{Eur. Phys. J. C \textbf{78}, no.1, 76 (2018)}.
\href{https://arxiv.org/abs/1709.02226}
{[arXiv:1709.02226]}

\bibitem{Zhong:2018exo}
T.~Zhong, Y.~Zhang, X.~G.~Wu, H.~B.~Fu and T.~Huang,
\textrm{The ratio $\mathcal {R}(D)$ and the $D$-meson distribution amplitude},
\href{https://doi.org/10.1140/epjc/s10052-018-6387-7}
{Eur. Phys. J. C \textbf{78}, no.11, 937 (2018)}.
\href{https://arxiv.org/abs/1807.03453}
{[arXiv:1807.03453]}

\bibitem{Zhang:2021wnv}
Y.~Zhang, T.~Zhong, H.~B.~Fu, W.~Cheng and X.~G.~Wu,
\textrm{$D_s$-meson leading-twist distribution amplitude within the QCD sum rules and its application to the $B_s\to D_s$ transition form factor},
\href{https://doi.org/10.1103/PhysRevD.103.114024}
{Phys. Rev. D \textbf{103}, no.11, 114024 (2021)}.
\href{https://arxiv.org/abs/2104.00180}
{[arXiv:2104.00180]}

\bibitem{Belyaev:1993wp}
V.~M.~Belyaev, A.~Khodjamirian and R.~Ruckl,
\textrm{QCD calculation of the $B\to\pi, K$ form-factors},
\href{https://doi.org/10.1007/BF01474633}
{Z. Phys. C \textbf{60}, 349-356 (1993)}.
\href{https://arxiv.org/abs/hep-ph/9305348}
{[arXiv:hep-ph/9305348]}

\bibitem{Colangelo:2000dp}
P.~Colangelo and A.~Khodjamirian,
\textrm{QCD sum rules, a modern perspective},
\href{https://doi.org/10.1142/9789812810458\_0033}
{[arXiv:hep-ph/0010175]}.

\bibitem{Narison:2014ska}
S.~Narison,
\textrm{Improved $f_{D^\ast_{(s)}}, f_{{B^\ast}_{(s)}}$ and $f_{B_{c}}$ from QCD Laplace sum rules},
\href{https://doi.org/10.1142/S0217751X1550116X}
{Int. J. Mod. Phys. A \textbf{30}, no.20, 1550116 (2015)}.
\href{https://arxiv.org/abs/1404.6642}
{[arXiv:1404.6642]}

\bibitem{Narison:2014wqa}
S.~Narison,
\textrm{Mini-review on QCD spectral sum rules},
\href{https://doi.org/10.1016/j.nuclphysbps.2015.01.041}
{Nucl. Part. Phys. Proc. \textbf{258-259}, 189-194 (2015)}.
\href{https://arxiv.org/abs/1409.8148}
{[arXiv:1409.8148]}

\end{thebibliography}
\end{document}